\documentclass[]{spie}  

\usepackage{amsmath,amsfonts,amssymb}
\usepackage[]{graphicx}
\usepackage[colorlinks=true, allcolors=blue]{hyperref}
\usepackage{hhline}
\usepackage{threeparttable}
\usepackage{rotating}

\title{A Radial Velocity Error Budget for Single-mode Fiber Doppler Spectrographs}

\author[a]{Andrew J. Bechter}
\author[a]{Eric B. Bechter}
\author[a]{Justin R. Crepp}
\author[b]{David King}
\author[a]{Jonathan Crass}

\affil[a]{Department of Physics, University of Notre Dame, 225 Nieuwland Science Hall, Notre Dame, IN, USA}
\affil[b]{Institute of Astronomy, University of Cambridge, Cambridge CB3 0HA, United Kingdom}

\authorinfo{Send correspondence to Andrew J. Bechter \\E-mail: abechter@nd.edu} 

\begin{document} 
\maketitle

\begin{abstract}
Single-mode fiber (SMF) spectrographs fed with adaptive optics (AO) offer a unique path for achieving extremely precise radial velocity (EPRV) measurements. We present a radial velocity (RV) error budget based on end-to-end numerical simulations of an instrument named “iLocater” that is being developed for the Large Binocular Telescope (LBT). Representing the first AO-fed, SMF spectrograph, iLocater's design is used to quantify and assess the relative advantages and drawbacks of precise Doppler time series measurements made at the diffraction limit. This framework can be applied for trade-study work to investigate the impact of instrument design decisions on systematic uncertainties encountered in the regime of sub-meter-per-second precision. We find that working at the diffraction-limit through the use of AO and SMF's allows for high spectral resolution and improved instrument stability at the expense of limiting magnitude and longer integration times. Large telescopes equipped with AO alleviates the primary challenges of SMF spectrographs. 
\end{abstract}

\section{Introduction}
\label{sec:RV}

Single-mode fiber (SMF) spectrographs offer a new technological avenue for achieving extremely precise radial velocity (EPRV) measurements for the detection of extra-solar planets \cite{schwab_12,robertson_12,crepp_14}. This type of spectrograph differs from conventional seeing-limited instruments by instead using adaptive optics (AO) to inject starlight directly into SMFs. Due to their small size (comparable to the wavelength of light), SMFs offer many advantages for EPRV measurements including: higher spectral resolution; improved optical quality; thermal stability and vacuum level (through a smaller opto-mechanical foot-print); the elimination of spatial modal noise; and negligibly small background levels. A diffraction-limited imaging input to SMF spectrographs also acts to decouple the instrument's optical design from the telescope diameter \cite{jovanovic_16}. Finally, the use of SMFs allows for seamless integration of existing photonic devices such as arrayed wave-guide gratings, fiber-based beam splitters, fiber based wave-front sensing, and fiber Fabry-P\'erot wavelength calibrators\cite{gurevich_14,schwab_15,harris_15,corrigan_16,jovanovic_17,anagnos_18}. 

Despite the obvious qualitative advantages to pursuing EPRV measurements with diffraction-limited instruments, rigorous quantitative RV uncertainty budgets in general are only now being fully developed and defined\cite{szenty_14,halverson_16}. In this paper we discuss the merits of pushing Doppler spectrographs to the diffraction-limit, with a particular focus on the RV impact of SMFs. Establishing a consistent methodology for computing each error budget term would be valuable for the development of new instruments, both in terms of determining achievable RV precision and with regard to the cost of design work. Currently, a number of terms have yet to be empirically validated, particularly those directly relating to the properties of SMFs. Therefore, given the promise of SMF spectrographs, developing error budgets for these distinct design deviations could be beneficial to future instruments.

\begin{sidewaysfigure*}[p]
	\centering
	\includegraphics[width = 9.2in]{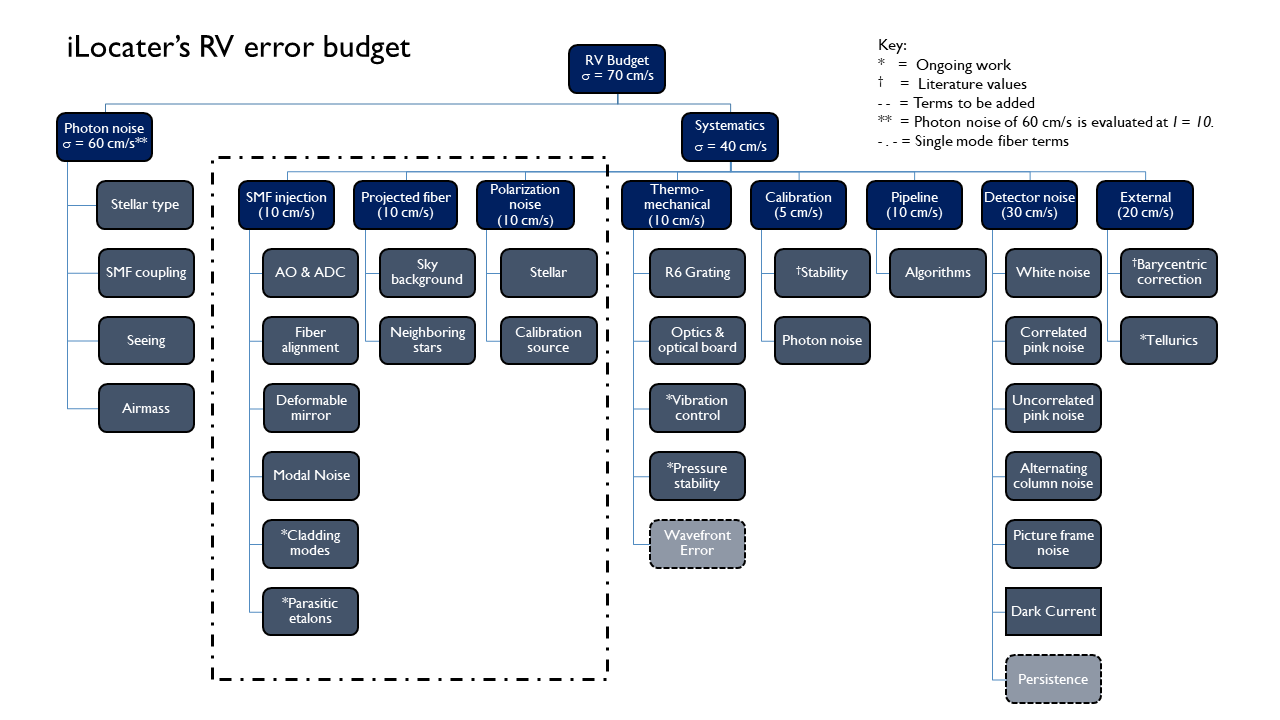}		
	\caption{Box diagram of the iLocater error budget. Gray boxes represent each term investigated, while blue boxes are the overarching category. Values listed in the blue boxes correspond to the allocated RV budget for each category. Terms grouped in the dashed box are identified as differing significantly from a MMF fed spectrograph. An asterisk (*) indicates the study is still on going whereas a dagger ($\dagger$) denotes a literature adopted value.Photon noise depends on stellar brightness and is calculated at I mag = 10. Light gray shaded boxes indicate new RV terms to be added in future work}
	\label{errbud}
\end{sidewaysfigure*}

We use the current design of a high-resolution, SMF, EPRV instrument called ``iLocater" as a benchmark system to generate a comprehensive RV error budget\footnote{Some aspects of iLocater's design have changed since this work was completed and therefore values should be treated as representative.}. iLocater is an $R = 150-240$k near infrared spectrograph being developed for the Large Binocular Telescope (LBT). The instrument is comprised of three major components: a SMF injection unit (called the 'Acquisition camera'); a diffraction-limited, Gaussian beam spectrograph housed in a thermally stable cryostat; and a vacuum spaced  Fabry-P\'erot calibration source \cite{bechter_16,crepp_16,crass_16,sturmer_17}. Each of these components has the ability to alter RV variations measured by the instrument and thus contribute to the overall error budget. Using this formalism, the impact of each design decision is assessed as an RV term that allows for hardware considerations to be related to scientific goals.

The error budget considers photon noise, instrument systematics, and external sources of error such as telluric lines from water vapor. While stellar variability is not quantified, SMF spectrographs offer the distinct possibility of measuring absorption line asymmetries in time, in an attempt to disentangle the effects of stellar noise from orbiting planets \cite{davis_17,rajpaul_17}. This science goal may be considered as inherent to the requirements of iLocater-like instruments. 

Figure~\ref{errbud} shows a comprehensive RV error budget for iLocater. Gray boxes represent each term considered individually, while blue boxes title the overarching category and list the allocated RV uncertainty . Detailed calculations of uncorrected RV values, and potential mitigation methods are provided for each term as a subsection. Terms marked with an asterisk (*) indicate the study is still ongoing whereas daggers ($\dagger$) indicate where values in the literature have been adopted. Much of this work was inspired by Szentgyorgyi et al. 2016, and Halverson et al. 2016 (Refs~\citenum{szenty_14,halverson_16}) with major differences being the use of SMFs instead of MMFs. In $\S$\ref{sec:Ph_Noise} we calculate the photon-noise-limited RV requirements and simulate the expected photon noise for various spectral types. In $\S$\ref{sec:SMF_inection}-$\S$\ref{sec:detector}, we analyze instrument systematics by assessing the RV uncertainty of individual contributors. A summary of all error terms is provided in $\S$\ref{sec:conclusion}.

\section[Photon Noise]{Photon Noise ($\sigma_{\rm ph}$)}
\label{sec:Ph_Noise}
This section presents instrument performance calculations based on the the stellar flux, signal-to-noise ratio, and photon noise limited RV uncertainties  ($\sigma_{\rm ph}$). These values are determined by creating an end to end (E2E) instrument throughput budget that includes Earth's atmosphere, existing optical coatings from the telescope, iLocater's acquisition camera and spectrograph components. All mirrors, lenses, dichroics, gratings, etc. are self-consistently incorporated into the analysis. A detailed investigation of SMF alignment and coupling efficiency is presented that includes both static (aberrations) and dynamic (atmospheric) effects. Fiber alignment tolerances and chromatic effects are quantified using semi-analytic methods in combination with modeling using Zemax.

\subsection{SNR requirement}
\label{sec:photon_noise}

\begin{figure*}[!t]
	\begin{center}
		\includegraphics[trim={2.5cm 0cm 2cm 0cm},clip,width=0.98\textwidth]{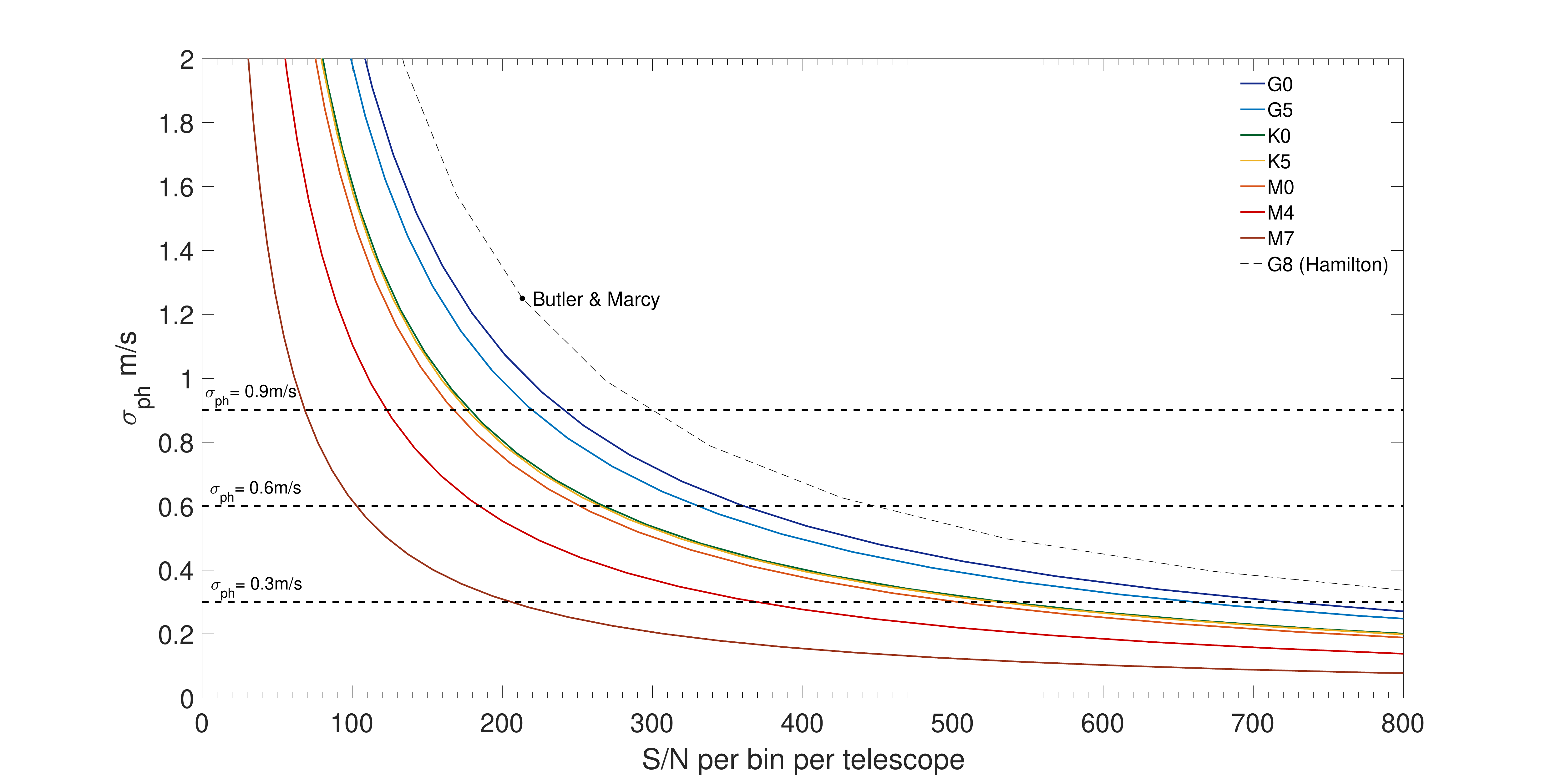} 
		\caption{Photon-noise limited RV precision as a function of SNR per spectrograph pixel. This plot is used to derive SNR requirements which can be used to derive an integration time and apparent magnitude for different stellar spectral types. Curves are generated assuming $v \sin{i}=2.5$ km/s with $R=180,000$ and 3 pixel per FWHM sampling across the YJ-bands. For reference, we over-plot an analysis of the Lick Hamilton spectrograph from Ref~\citenum{butler_96}. This curve was used to verify photon-noise estimates from which we find agreement to within 10\%.}
		\label{fig:SNR}
	\end{center}
\end{figure*}

To determine required flux, we model the transmission of a star as its spectrum would be sampled by the spectrograph detector using the formalism from Ref. \citenum{butler_96} for calculating the Photon noise-limited RV uncertainty $\sigma_{\rm ph}$:
\begin{equation}
\sigma_{\rm ph}=\frac{1}{\sqrt{\sum \left(\frac{dI/dV}{\epsilon_I} \right)^2}}
\end{equation}
where $dI/dV$ represents the slope of the measured stellar intensity as a function of wavelength (expressed in velocity units) and $\epsilon_I=\sqrt{N_{\rm ph}}/N_{\rm ph}$ is the fractional Poisson error. Figure~\ref{fig:SNR} shows results for photon-noise limited RV uncertainty (single measurement precision) as a function of SNR for different spectral types. This exercise establishes the required SNR which in turn corresponds to a flux level in photons per second. Given required SNR values, we then determine a lower limit for incident light to the spectrograph, including effects due to SMF coupling efficiency from iLocater's acquisition camera as a function of seeing conditions, airmass, and stellar parameters (spectral type, apparent magnitude, $v \sin{i}$).  

Allocating $\sigma_{\rm ph}=0.6$ m s$^{-1}$ RV uncertainty from photon noise, the required SNR=208 per pixel for an M4V star (Table~\ref{tab:SNR}) where
\begin{equation}
\mbox{SNR}=\sqrt{N_{\rm ph}},
\end{equation}
and $N_{\rm ph}$ represents the total number of photons detected. SNR requirements for other $\sigma_{\rm ph}$ values shown in Figure~\ref{fig:SNR} are also listed in Table~\ref{tab:SNR}.Later spectral types contain intrinsically higher Doppler content due to the depth and number density of absorption lines in the YJ-bands and therefore require lower SNR for a given level of photon noise. SNR values in parenthesis represent the increased SNR required when considering the reduced spectral band resulting from telluric line contamination. This value should be considered a worst case scenario with zero telluric correction. 

\begin{table}[!t]
	\begin{threeparttable}
		\centerline{
			\begin{tabular}{|c|ccccccc|}
				\hhline{|========|}
				\multicolumn{8}{|c|}{SNR Requirements} \\
				\hhline{|========|}
				RV (m/s) &G0V &G5V &K0V & K5V & M0V & M4V & M7V \\
				\hhline{|=|=======|}
				$\sigma_{\rm ph}$ = 0.9 & 275 (440) & 254(400)& 206 (325)& 204 (330)& 195 (300) & 138 (220)& 77 (124)\\
				$\sigma_{\rm ph}$ = 0.6 & 414 (660)& 379 (605)& 309 (490)& 306 (493)& 291 (460) & 208 (330)& 115 (184)\\
				$\sigma_{\rm ph}$ = 0.3 & 826 (1300)& 757 (1200)& 618 (980)& 613 (987)& 503 (925) & 415 (660)& 231 (370)\\
				\hline
		\end{tabular}}
		\label{tab:SNR}
	\end{threeparttable}
\end{table}

\subsection{Throughput}
\label{sec:tput}
Modeling an accurate throughput is essential for calculating the achievable SNR, and limiting magnitude. The total integrated throughput of the spectrograph, $T_{spec}$, is represented by:
\begin{equation}
\label{eqn:spec_tput}
T_{\rm spec} = \int \: t_{\rm tel}(\lambda) \: t_{\rm ent}(\lambda) \: t_{\rm LBTI}(\lambda) \: t_{\rm acq}(\lambda) \: \alpha(\lambda) \: t_{\rm fl}(\lambda) \: t_{\rm sp}(\lambda) \: d\lambda
\end{equation}
where the various subscripted $t(\lambda)$ terms, represent the wavelength-dependent throughput (0-1) of: the telescope mirrors, the wavefront sensor (WFS) pickoff (entrance window), the large binocular telescope interferometer (LBTI), iLocater's acquisition camera, the fiber link, and spectrograph respectively. All of these surfaces are modeled using commercial optical coatings. Out of the throughput terms above, the only dynamical term is SMF coupling efficiency, $\alpha(\lambda)$, which is proportional to the Strehl ratio delivered by the AO system\cite{wagner_82}. This term requires care to calculate since the Strehl ratio is dependent on stellar apparent magnitude and the observational conditions, 
\begin{equation}\label{eqn:eta_full}
\alpha(\lambda)=\alpha(\lambda,\theta,\sec{z}),
\end{equation}
where $\theta$ represents zenith seeing and $\sec(z)$ is airmass. We estimate SMF coupling by calculating three individuals terms, asserting that they are separable,

\begin{equation}
\alpha(\lambda,\theta,\sec{z})= S_{\rm AO}(\lambda,\theta,\sec{z}) \: S_{\rm static}(\lambda) \: r(\lambda)
\label{eqn:coupling}
\end{equation}
where $S_{\rm AO}(\lambda,\theta,\sec{z}) < 1$ is a dynamic term corresponding to the Strehl ratio delivered from the AO system, $S_{\rm static}(\lambda) < 1$ represents Strehl ratio degradation due to alignment errors between the incident beam and the fiber tip, and $r(\lambda) < 1$ represents Fresnel reflection losses at both ends of the fiber, for which we conservatively assume $r=0.96^2$ for all wavelengths.

SMF coupling efficiency can be calculated using the overlap integral of the fundamental fiber mode and telescope pupil as described in Ref~\citenum{ruilier_98}, although this method requires an exact knowledge of the electric field. We use this formalism to calculate the theoretical coupling efficiency of the fiber and telescope PSF,$S_{\rm static}(\lambda)$, which includes static aberrations and expected fiber misalignment. An alternative method discussed in Ref~\citenum{wagner_82} shows that SMF coupling efficiency is approximately equal to the Strehl ratio which is modeled by scaling results from Ref \citenum{pinna_16} depending on the seeing and airmass of the simulated target star. We use this method to estimate fiber coupling losses due to AO correction, $S_{\rm AO}(\lambda,\theta,\sec{z})$. Details of this term will be presented in Bechter E. et al. 2018 (a), in prep. 

\begin{figure}[p]
	\centering
	\includegraphics[width=5.5in]{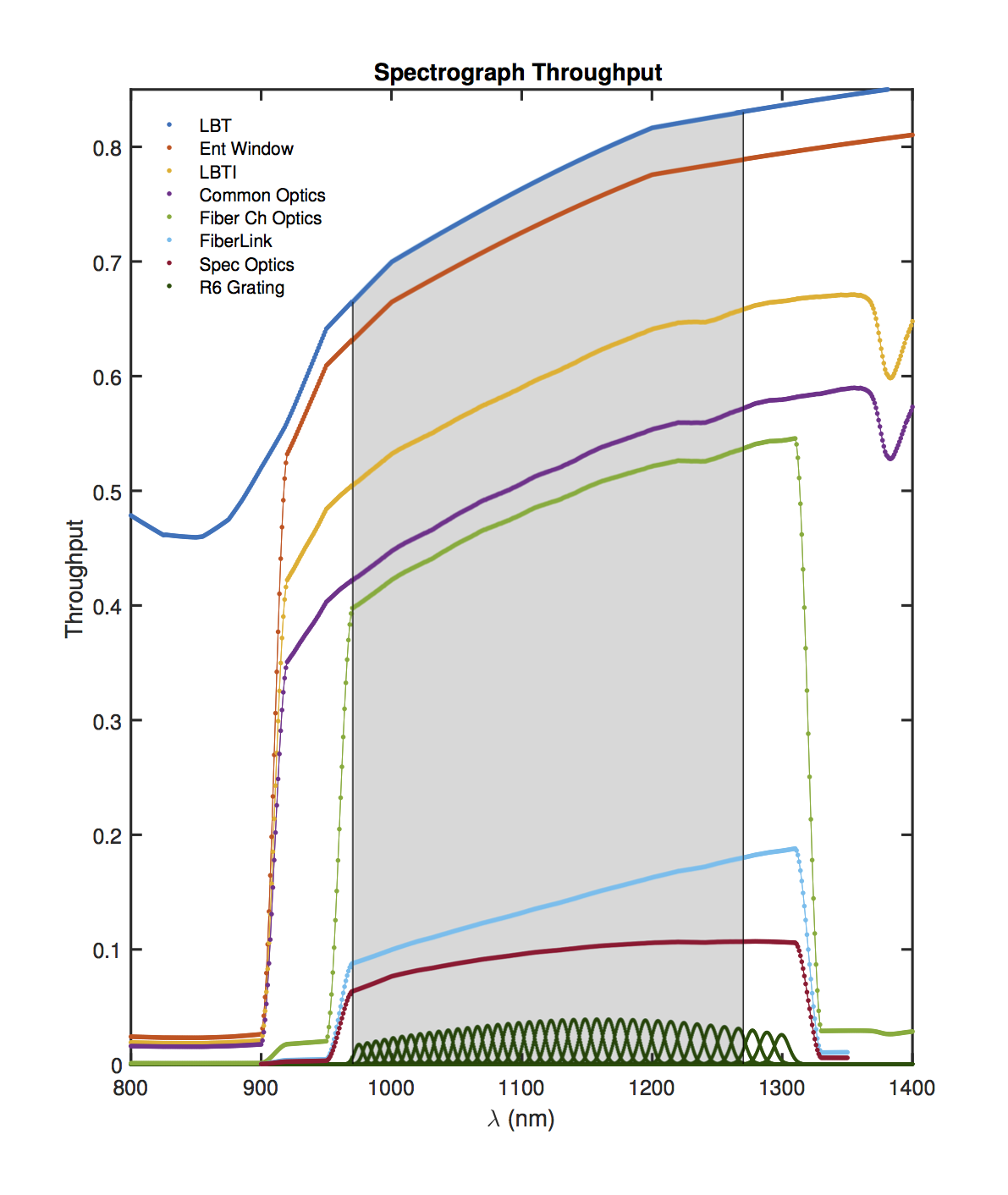}
	\caption{Wavelength dependent instrument throughput. Each curve shows the progressive throughput from one sub-section of the instrument to the next. Transition regions occur on either side of the science band (shaded) to ensure maximum throughput is delivered to the spectrograph. The FiberLink term is dominated by the Strehl ratio delivered by the AO system and contributes the single largest loss. Injection losses create a slight ``bend" in the transmission curve due to the chromatic coupling losses introduced by the ADC and triplet lenses, however the R6.1 grating efficiency dominates the final curve shape, creating a peak throughput near the middle of the bandpass. The Integrated transmission from the LBT primary mirror to the fiber tip is $\sim$~47.3\% over the band $\lambda=0.971-1.270 \; \mu$m and the peak efficiency of each order is $\sim$~1.5-2\%. }
	\label{fig:Tput}
\end{figure}

The results of these calculations can be found in Figure~\ref{fig:Tput} which shows the E2E throughput for typical AO performance and bright guide stars is approximately 2$\%$. In this figure, collective surfaces are grouped together to reduce the number of curves and highlight key components. For reference, the LBT consists of the primary, secondary, and tertiary UV-enhanced aluminum coated mirrors. The entrance window and LBTI consists of a new entrance window dichroic designed specifically for iLocater (this reflects visible light to the WFS and transmits NIR light to iLocater), and gold coated elliptical, pupil and roof mirrors contained within LBTI. The coatings of the optics within the acquisition camera (common optics and fiber channel) are modeled as commerical gold and NIR lens coatings. Throughput losses are dominated by the AO fiber coupling term (contained within the FiberLink term in light green), the R6 grating (dark green curve), and LBT mirror coatings (blue curve). To significantly improve throughput, higher efficiency gratings and better AO performance, would need to be implemented. 


\subsection{Simulated SNR}
\label{sec:SNR}

Following light from the star to detector, we simulate the SNR using a flux preserving algorithm to map counts from each order into detector pixel bins. Simulated detector frames are created by scaling a stellar spectrum to the desired apparent magnitude, multiplying by atmospheric and instrument throughput (\S~\ref{sec:tput}), and then finally clipping (using a flux preserving binning algorithm) the resulting spectrum down onto a detector grid which is sampled according to iLocater's optical model. These frames are read into the pipeline where 2D orders are extracted and collapsed into a 1D spectrum. The counts in each bin of the 1D spectrum are used to calculate SNR.   

\begin{figure}
	\centering
	\includegraphics[width=5.4in]{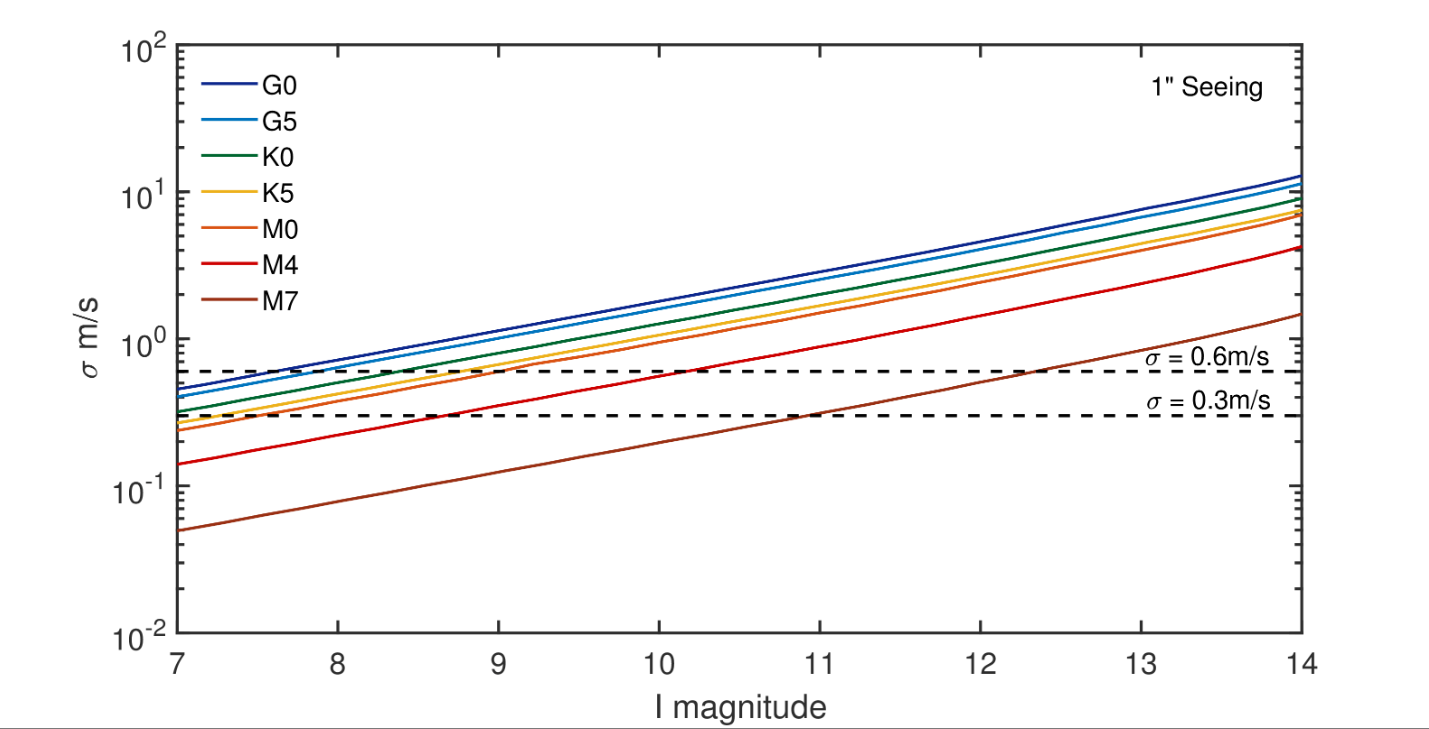}
	\caption{Photon noise as a function of I magnitude in 1'' seeing. Mid-late M-stars meet the $\sigma_{\rm ph}=0.6$ m/s requirement at $I \leq 10$. Earlier type stars require higher SNR to make up for less Doppler information available in the YJ-bands and thus have brighter limiting magnitudes for a given precision goal.}
	\label{fig:sigma_1}
\end{figure}

Figure \ref{fig:sigma_1} shows iLocater's photon noise limited precision as a function of apparent magnitude for $\theta = 1$". iLocater's limiting magnitudes for spectroscopy are assessed at $\sigma_{ph}=0.6$ m/s, and $\sigma_{ph}=0.3$ m/s. In these simulations we consider the worst case scenario of no telluric correction, resulting in wholesale removal of stellar lines with any overlapping sky absorption with an exposure time of $\Delta t$ = 30 minutes.

Photon-noise-limited RV precision results for mid-type M-stars with $\sigma_{\rm ph}$ = 0.6 m/s can be achieved for $I~<~9$. Earlier type stars of the same $I$-magnitude will require more observation time or better seeing conditions to achieve the same precision. For brighter targets, $I~<~7.5$, $\sigma_{\rm ph}<0.6$ m/s can be achieved essentially irrespective of spectral type. An example of such a star is 61 Virgo: a G6.5V type star with an $I$ magnitude of 4.1 and multiple documented exoplanet detections \cite{vogt_10}. By scaling the flux from the left side of Figure~\ref{fig:sigma_1} according to the magnitude of 61 Virgo and extending results from Figure~\ref{fig:SNR}, we conservatively calculate $\sigma_{\rm ph}$ = 0.2-0.3 m/s. At the faint end, iLocater is primarily limited by the Strehl ratio delivered by the LBT AO system. For these cases it may be necessary to integrate for longer than $\Delta t = 30$ minutes depending on target properties.

We emphasize that the properties of a SMF benefit spectrograph stability and eliminate certain sources of systematic noise. As a consequence however, the use of a SMF places more stringent requirements on photon noise, which is governed by AO performance. 

\section[Instrument Systematics]{Instrument Systematics ($\sigma_{\rm sys}$)}
Systematic RV terms are calculated by numerically simulating spectra at nominal conditions and cross correlating with a spectrum in a perturbed state. Spectra are simulated using iLocater's end to end (E2E) instrument simulator, detailed in Bechter E. et al. 2018 (b), in prep, and extracted from the simulated detector to include physical sampling effects of the detector. We group systematic RV uncertainties related to the intrinsic properties of SMFs into three sections: SMF injection errors ($\S$\ref{sec:SMF_inection}), sky background ($\S$\ref{sec:projectedfiber}), and polarization noise ($\S$\ref{sec:polarization}). Thermo-mechanical terms ($\S$\ref{sec:thermal}) consider spectrograph optics and optical board stability with specific attention to material selection. Additional consideration is given to the wavelength calibration source ($\S$\ref{sec:wavecal}), tellurics ($\S$\ref{sec:tellurics}), Barycentric Correction ($\S$\ref{sec:bary}), the data reduction pipeline ($\S$\ref{sec:software}), and simulations of H4RG detector noise ($\S$\ref{sec:detector}).  

\subsection[SMF Injection]{SMF Injection ($\sigma_{\rm fiber}$)}
\label{sec:SMF_inection}
Efficient SMF coupling relies on active optical beam correction using AO. In the monochromatic case, fiber coupling errors result in a loss of photons and do not necessarily introduce systematic RV terms. However, losses in fiber coupling due to residual dispersion, AO correction, and fiber alignment are wavelength dependent and changes in the observing conditions (seeing, airmass) from one observation to another varies the chromatic coupling efficiency. Thus, the same star observed under differing nightly conditions will have a variable spectral gradient with a quantifiable RV shift. In principal, this effect occurs when using both single-mode and multi-mode fibers, the difference between them primarily being the magnitude of chromatic loss\cite{halverson_16}. We identify the AO system, ADC, and fiber alignment as the worst offending components that cause chromatic coupling variations\footnote{The RV effects due to anisotropic dichroic coatings and degradation of mirror coatings is an on-going study and has not yet been quantified.}. These effects are quantified by simulating the chromatic coupling efficiency under variable observing conditions and measuring the RV offset by cross correlation. A breakdown of the above errors are treated in detail in \S\ref{sec:ADC_AO} and \S\ref{sec:fiber}. 

Due to the fundamental guiding properties of single-mode wave-guides, only a portion of light incident on the fiber face will remain bound within or near to the core region. Any remaining light is lost due to Fresnel reflection, radiation modes or cladding modes. While radiation modes lead to a loss of photons, cladding modes can result in undesirable effects that degrade RV precision. In addition, back reflections caused by Fresnel reflection at fiber end-end connections (as used in the iLocater FiberLink), can create cavities within the fiber itself, or in air gaps between the fiber end faces resulting in an etalon effect. Cladding modes and undesirable etalons are discussed in $\S$\ref{sec:cladding_modes} and $\S$\ref{sec:paretalons}.

\subsubsection[AO \& ADC]{Adaptive Optics \& Atmospheric Dispersion Correction ($\sigma_{\rm AO}$)}
\label{sec:ADC_AO}
Efficient SMF injection requires ``extreme" AO correction and exquisite dispersion correction. iLocater's ADC design minimizes dispersion effects by using a triplet prism design that maintains precise alignment throughout the course of an observation using a zero-beam deviation optical design \cite{kopon_13}. Nevertheless, the tolerances imposed through the use of a SMF are small ($\sim\rm1\mu$m) and residual ADC effects can contribute to iLocater's error budget. We simulate the impact of residual atmospheric dispersion using iLocater's acquisition camera optical design, with the sky representing the ``first surface" of optical components, and model the influence of flux gradients on RV precision using iLocater's E2E data reduction pipeline. Although performance of the ADC and AO system are coupled, we first investigate them separately to understand the influence of each component.

\begin{figure}
	\centering
	\includegraphics[width = \textwidth]{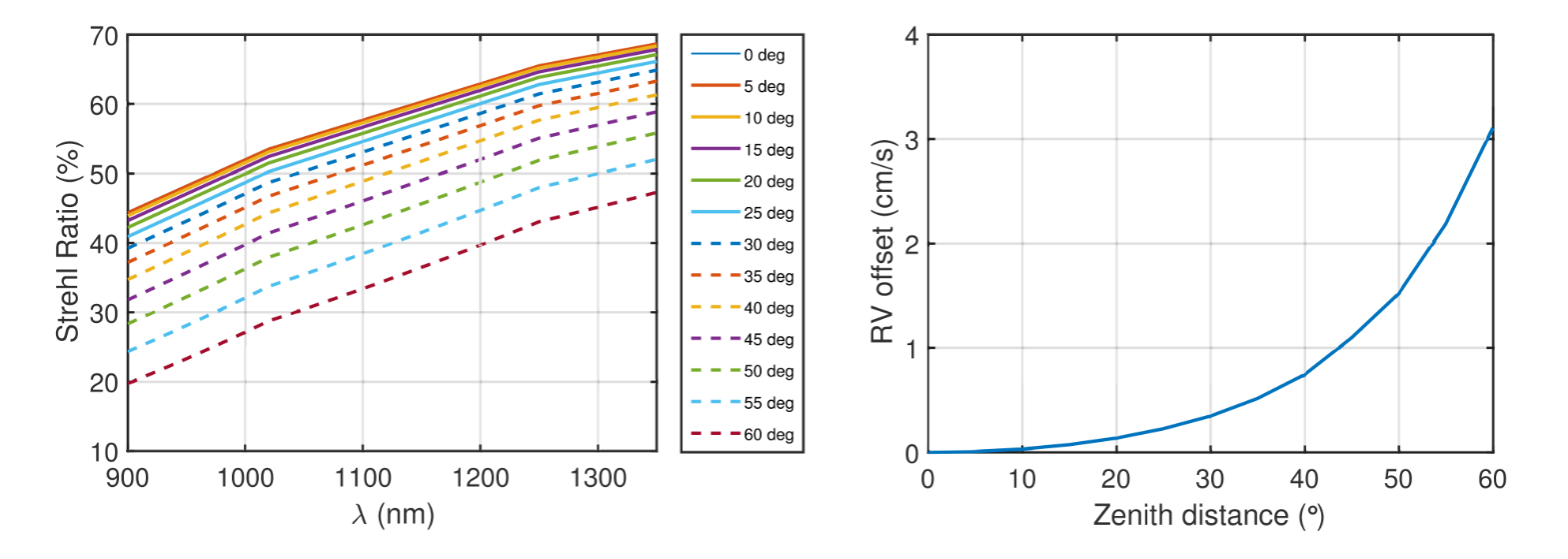}	
	\caption{\textbf{Left}: Strehl ratio as a function of wavelength for zenith angles 0-60 degrees (M0V star at I mag~=~9 in 1'' seeing). Fiber coupling efficiency directly follows the Strehl ratio. A slight change in slope with airmass can induce RV shifts. \textbf{Right}: RV precision as a function of zenith angle, where the RV offsets are measured relaitve to z = 0. The offset follows a smooth function similar to the changing Strehl ratio.}
	\label{fig:SR}
\end{figure}

When estimating an average fiber coupling efficiency, the Strehl ratio delivered by the AO system is clearly the dominant term. The resulting spectrum coupled into the SMF is approximated by the product between the stellar spectrum and the Strehl ratio curve with only minor losses on average from residual dispersion. To approximate Strehl degradation with increasing airmass, we incorporate the zenith distance, $z$, analytically: 
\begin{equation}\label{eqn:airmass}
S(\lambda,\theta,z) = \exp(-|\log_{e}(S_0(\lambda,\theta))|\sec{z}),
\end{equation}
where $S_0(\lambda)$ represents the Strehl ratio at $\sec{z}=1$. The result is a changing chromatic throughput depending on the airmass. Figure~\ref{fig:SR} shows the Strehl ratio as a function of wavelength, in steps of $\Delta z = 5$ degrees. At first glance the Strehl ratio appears to shift downwards by a fixed percentage however, closer inspection reveals there is a changing gradient (20\% change on the red end compared to 25\% on blue end). This subtle change introduces an RV shift. Radial velocity uncertainties are calculated by cross correlating the fiber coupled spectra at each different airmass with a template spectrum with an airmass of 1. The calculated velocity offset follows a smooth function similar to the changing Strehl ratio, with a maximum value of 3~cm/s.

\begin{figure}
	\centering
	\includegraphics[width = \textwidth]{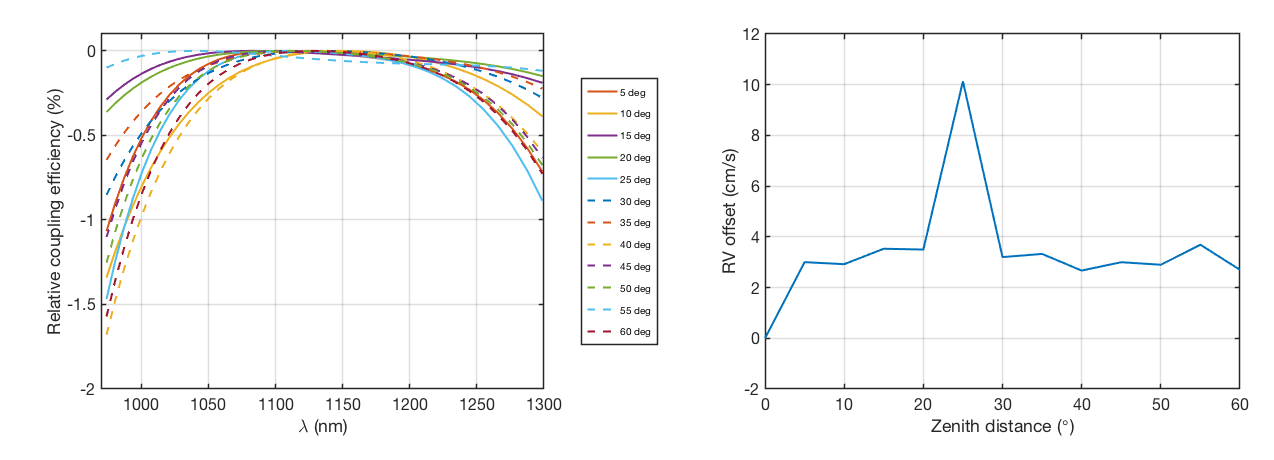}	
	\caption{(\textbf{Left}:) Fiber coupling loss due to residual dispersion after the ADC as a function of wavelength for zenith angles $\sec{z}=0-60$ degrees (M0V star at $I\approx9$). As opposed to the Strehl ratio shown in Figure~\ref{fig:SR}, residual dispersion has little trend with airmass on account of a well-designed ADC. (\textbf{Right}:) RV precision as a function of zenith angle. The RV offset is somewhat chaotic as there is no clear trend with airmass} 
	\label{fig:ADC}
\end{figure}

Uncorrected residual dispersion from the ADC will compound the chromatic losses due to the AO system. We first disperse the beam according to the ADC performance calculated in Zemax from $\sec{z}=5-60$ degrees. Coupling efficiency for each wavelength is calculated using the simulated radial dispersion relative to the central wavelength, shown in Figure~\ref{fig:ADC} (left panel), where a perfect ADC would produce lines with zero slope. In practice, residual dispersion in the fiber focal plane follows a second order polynomial and is limited to less than 10\% of the fiber diameter. In a similar way to the AO test, the RV offsets are calculated by cross correlating each airmass from 5-60 degrees with a template spectrum at an airmass of 1 (Figure~\ref{fig:ADC}, right panel). The maximum fiber coupling loss relative to the central wavelength is ~2$\%$ which results in an RV offset of a few cm/s, a similar contribution to that of the AO system.   

\begin{figure}
	\centering
	\includegraphics[width =\textwidth]{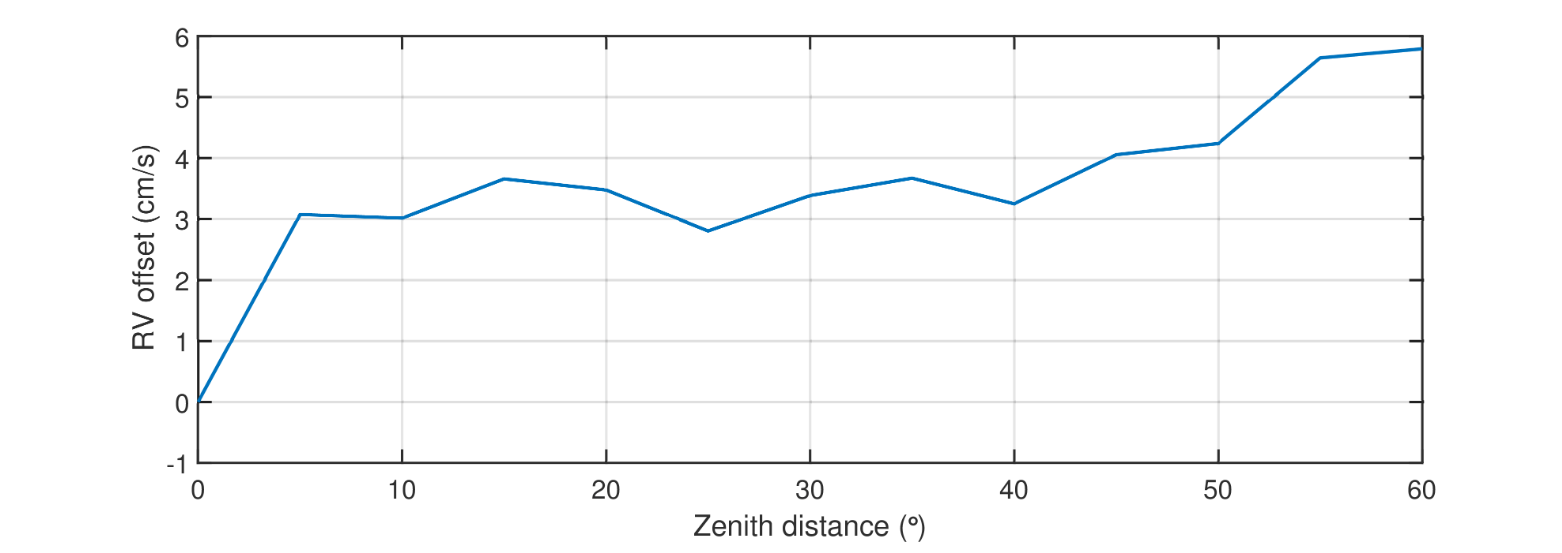}	
	\caption{Combined effect of AO system and ADC RV uncertainty as a function of airmass.}
	\label{fig:AO_ADC}
\end{figure}

As both the AO system and ADC performance depend on airmass, the Strehl ratio and residual dispersion should be simultaneously degraded accordingly. The combined error due to the AO system and the ADC is shown in Figure~\ref{fig:AO_ADC}. These effects might be mitigated in post-processing through continuum normalization; however, until such algorithms are validated, we conservatively use an estimate of 6 cm/s for the combined AO and ADC correction. 

\subsubsection[Fiber Alignment]{Fiber Alignment ($\sigma_{\rm Fiber}$)}
\label{sec:fiber}
\begin{figure}
	\centering
	\includegraphics[width =\textwidth]{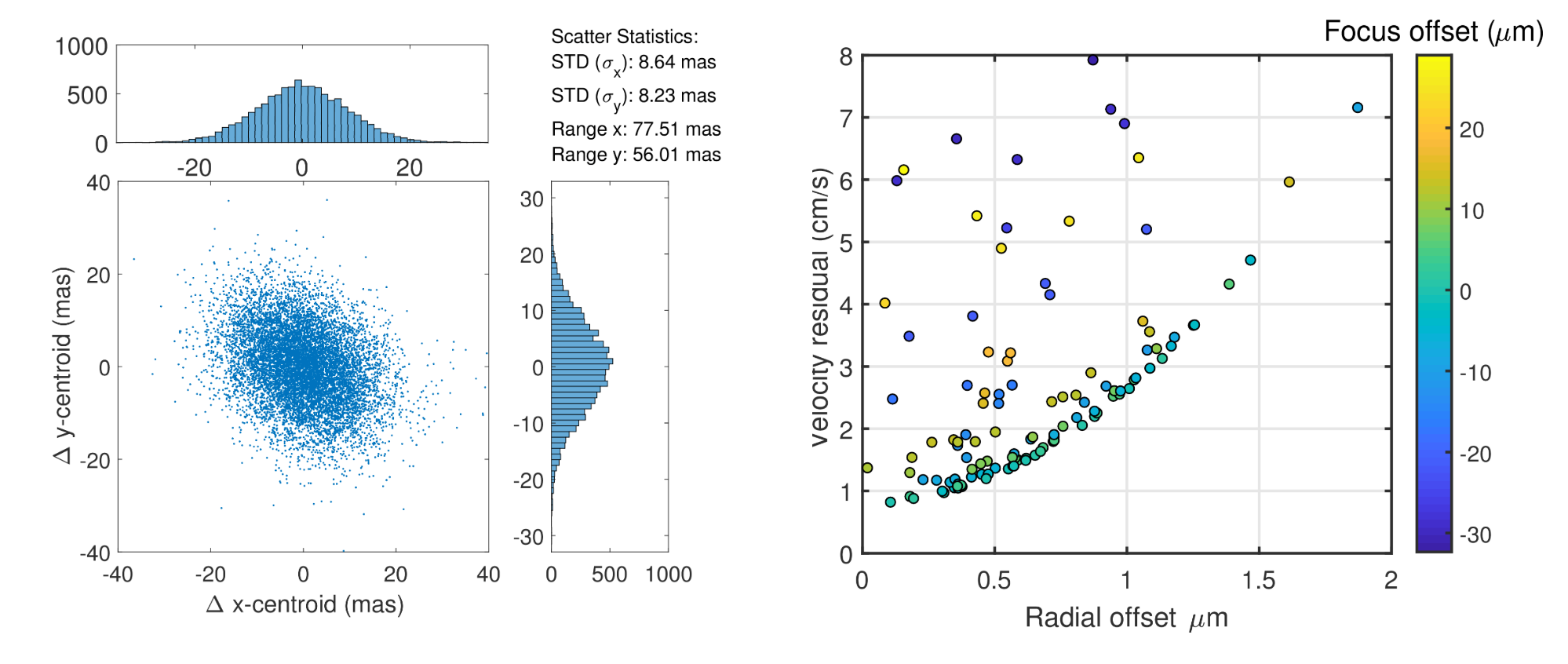}	
	\caption{(\textbf{Left}:) On-sky scatter of the stellar PSF with histograms showing the distribution of centroid motion. Residual tip/tilt motion is calculated by measuring the relative PSF distance from the centroid mean and converting the pixel scale to mas. Centroid scatter is a visual representation of PSF motion over time. (\textbf{Right}:) Velocity offsets as a function of 3D fiber position. Radial offsets relative to the injected AO beam are calculated in the x-y plane of the fiber. Focus offsets (z-dimension) are represented by the color scale. Increasing both focus errors and radial offsets introduce RV offsets at the few cm/s level.}
	\label{fig:fiber_rv}
\end{figure}

Misalignments of the fiber and input beam introduce variations in the chromatic coupling efficiency. For instance, a fiber misaligned by 1 $\mu$m will cause greater loss at shorter wavelengths than longer wavelengths following the diffraction-limited size of the PSF. Residual tip/tilt errors not only reduce coupling efficiency, but also limit the precision of fiber alignment. To account for fiber misalignments we use on-sky telescope data shown in Figure~\ref{fig:fiber_rv} (left panel), which has a $1\sigma$ value of $\sim$ 9~mas. We estimate that scatter will be improved by a factor of three due to iLocater's internal beam stabilization loop (comprising a fast steering mirror (FSM) and quad-cell), thus reducing the $1\sigma$ value to 3~mas. 

Given the measured scatter, we take a statistical approach to estimate the fiber's radial position by performing a Monte-Carlo simulation. We randomly draw a radial fiber position from a Gaussian distribution, setting $\sigma$ equal to the PSF scatter. For each iteration, the slope in fiber coupling efficiency varies the instrument throughput curve. RV uncertainties for each new three dimensional fiber position, which are calculated relative to a spectrum produced by a perfectly positioned fiber, are shown in Figure~\ref{fig:fiber_rv}. A visible trend can be seen with increasing radial positioning errors shown in light blue and green data points. For a fiber with little defocus error, the RV offset is primarily $\leq$ 5~cm/s. Including large focus offsets (purple and yellow points) the uncertainty increases to 8~cm/s.

\subsubsection[DM Doppler Shifts]{Deformable Mirror Induced Doppler Shifts ($\sigma_{\rm DM}$)}
\label{sec:DMshifts}
Piston from the AO system's deformable mirror can induce Doppler shifts resulting from its bulk motion. We place an upper-limit on this effect by noting that typical AO systems have a stroke of several waves ($\sim 10\mu$m). The fastest possible induced motion (1~ms) thus corresponds to a RV of $\approx 10 \; \mu m \; \mbox{ms}^{-1}$, or $\sigma_{\rm DM}=1$ cm s$^{-1}$. In practice, the uncertainty from this effect is much smaller since the integration time averages over many AO loop cycles. 

\subsubsection[Modal Noise]{Modal Noise ($\sigma_{\rm modal}$)}
\label{sec:modal}
``Modal noise" refers to the interference pattern created when multiple spatial modes supported by an optical wave-guide are imaged by the spectrograph. As fiber boundary conditions change over time due to temperature and pressure changes or other mechanical effects, the interference pattern changes creating a spurious Doppler shift. Existing seeing-limited RV spectrographs must mix the various spatial modes by modulating the fiber (mechanical agitation) while also scrambling its input signal through the use of octagonal fibers or other means \cite{fischer_16}. iLocater will use SMF's to eliminate the effects of spatial modal noise entirely. SMF's simultaneously offer high spectral resolution and a small opto-mechanical footprint. Combined, these effects served as the original motivation for building the instrument. 

\subsubsection[Cladding modes]{Cladding Modes ($\sigma_{\rm clad}$)}
\label{sec:cladding_modes}

The cladding in a SMF is surrounded by another medium called the coating. If the coating material has a lower index of refraction than the cladding, total internal reflection can occur. The result is so-called cladding modes, the intensity distribution of which is not restricted to the region in or immediately around the fiber core. Although these modes are more highly attenuated than the fundamental mode, they can persist in the cladding and exit the fiber into the spectrograph if not stripped properly. In addition, cladding modes exit the fiber with a smaller NA, and manifest in the focal plane as a speckle pattern that is superimposed on the fundamental mode and suffers from modal noise\cite{melnikov_96}.



Improperly injected starlight increases the fraction of power that excites cladding modes. Depending on the properties of the surrounding coating, cladding modes may either propagate over long distances or may be strongly attenuated by leakage into the coating. The latter situation is common, particularly for SMFs. iLocater's fibers are coated with a layer of Dual Acrylate ($\approx$100 $\mu$m thick) with refractive index $n=1.482$ whereas the pure silica cladding is $n=1.4507$ at $\lambda = 980$~nm, thus prohibiting total internal reflection. 
Coupling of power between core and cladding modes is not typically a problem due to the difference in propagation constants between the two regions. Leakage can occur from one to the other however, if the regular structure of the fiber is perturbed (such as the periodic disturbance utilized in fiber Bragg gratings). No such perturbations are expected in iLocater's fiberLink so we do not anticipate significant leakage from cladding to core or visa versa.


In the case of short lengths of fiber, light in the cladding may not be completely attenuated if the injection parameters of the fiber are poorly matched. As a default, iLocater will wrap the fiber end as close to the injection point as possible to promote further attenuation of cladding modes in the spatial transient region \cite{snyder_83}. Splicing the short length to a long segment of fiber will also help attenuate any cladding modes by allowing more propagation length. iLocater will have a short (2m) length of fiber to couple starlight into which is connected to a long (40m) fiber which goes to the spectrograph. If additional attenuation is required, a drop of index-matching fluid can be used on the fiber to completely eliminate persistent modes. This method should only be utilized if completely necessary as it requires the protective coating material to be removed, leaving the fiber more vulnerable to damage.

\subsubsection[Parasitic Etalons]{Parasitic Etalons ($\sigma_{\rm etalon}$)}
\label{sec:paretalons}
iLocater's fiber-link consists of four segments of fiber. Three of the segments are short, ranging from 2-7 meters and one is much longer at 40 meters. Each segment is connected using 8 degree angle polished AVIM connectors which allow for extremely high throughput. If there is a refractive index change at each end of a fiber, such as an air-glass interface, Fresnel reflection will cause light to back propagate through the fiber and form an etalon. Such effects could, in theory, imprint an undesirable transmission function on a coherent spectrum such as the calibration source. To ensure there is no such effect, the coherence length of the cavity must be significantly longer than that of the light source. 

The output of iLocater's Fabry-P\'erot etalon is has a Lorentzian optical spectrum with a coherence length of:
\begin{equation}
L = \frac{c\pi}{\delta\nu}
\end{equation}
where $\delta\nu$ is the line width measured at the FWHM. This coherence length is the propagation length after which the magnitude of the coherence function has dropped to the value of $1/e$. For iLocater, the coherence length of the 10GHz spectrum is $L\sim$10mm, which is much shorter than the length of any individual fiber segment. In addition to providing excellent throughput, the angle polished AVIM connectors are precisely aligned and polished to be in contact with each other, eliminating potential air gaps.

\subsection[Projected Fiber Angle]{Projected Fiber Angle ($\sigma_{\rm angle}$)}
\label{sec:projectedfiber}
A SMF subtends a very small angle on the sky. As the collecting area of the fiber is designed to roughly match the diffraction limited size of a point source, very little off axis background light is collected. The projected diameter of iLocater's SMF on the sky is $\sim42$~mas, which is more than an order of magnitude smaller than a typical diameter of a multi-mode fiber used in astronomy ($\sim1$~arcsec). Furthermore, the collecting area is $\sim1\times10^{-3}$~arcsec$^2$, almost three orders of magnitude smaller (0.2\%) than a 1~arcsec$^2$ fiber. In the following sections we consider the impact of sky background and neighboring star contamination on RV precision. 

\subsubsection[Sky Background]{Sky Background ($\sigma_{\rm OH}$)}
\label{sec:skyback}
The $^*$OH radical hydroxyl is created in Earth's atmosphere at a layer of $\approx$ 6-10 km elevation when water molecules are dissociated by sunlight, cosmic rays, and other effects. As the molecules recombine in a reaction between atomic hydrogen and ozone, rotation-vibration modes from $^*$OH emit light at numerous discrete wavelengths between $\lambda = 0.61-2.62 \mu$m, creating bright emission features \cite{meinel_50}. This effect, often referred to as ``air-glow," can create significant SNR issues with seeing-limited spectrographs that observe faint astronomical objects, especially at low resolution. Further, the intensity of OH-lines can change over time requiring careful calibration. 

We simulate sky background from raw sky emission files available from the Gemini observatory.\footnote{http://www.gemini.edu} The original files are generated from sky transmission files \cite{Lord_92}, subtracted from unity to give an emissivity, and then multiplied by a blackbody function of temperature $T=273$ K. Added to these are: an OH emission spectrum, O$_{2}$ lines, and the dark sky continuum, approximated as a 5800 K gray body multiplied by the atmospheric transmission and scaled to produce 18.2~mag/arcsec$^2$ in the H band, as observed on Mauna Kea by Ref~\citenum{Maihara_93}. 

\begin{figure}
	\centering
	\includegraphics[width = 6.5in]{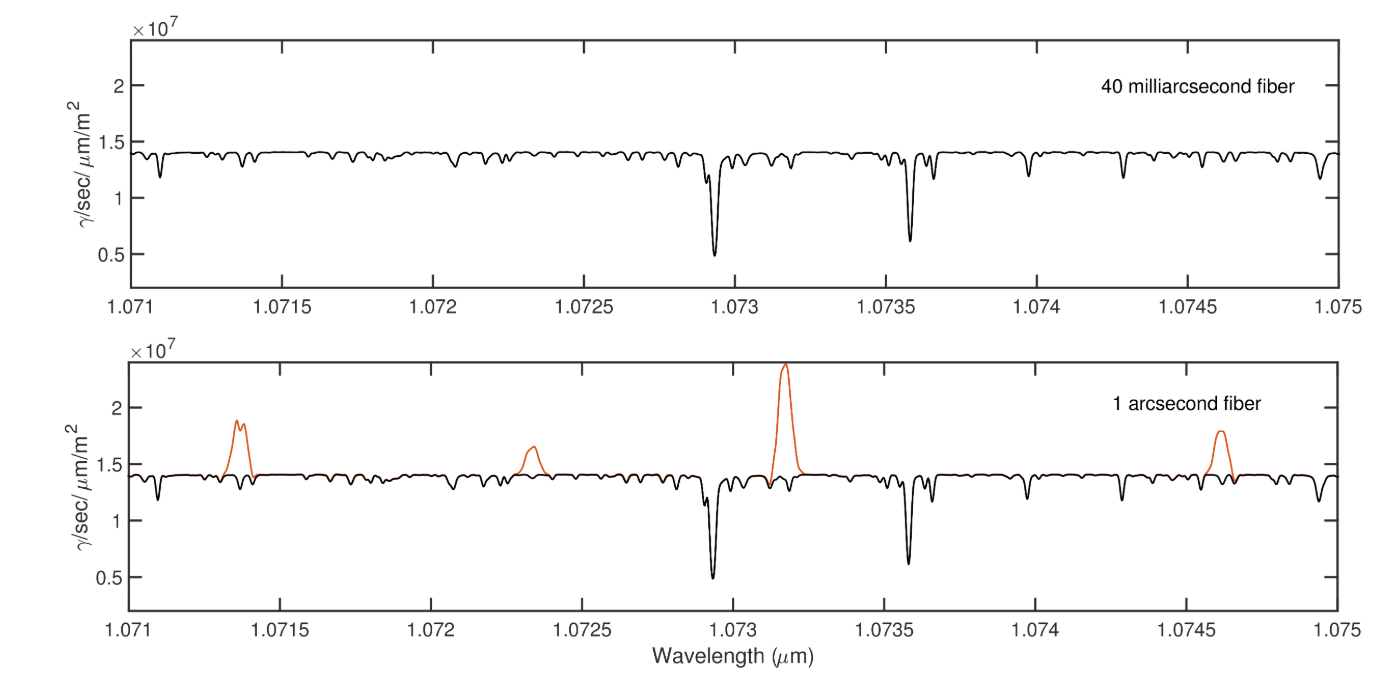}	
	\caption{(\textbf{Top}:) An M0V spectrum with added sky background collected by a SMF (orange), is over-plotted onto a stellar spectrum with no background (black). The fractional contribution due to sky background is indistinguishable by eye but can be measured at the cm/s level via cross correlation. (\textbf{Bottom}:) Using the same color scheme as the top panel, an M0V spectrum with added sky background collected by a 1" multi-mode fiber is over-plotted on a stellar spectrum with no background. Large OH lines are immediately visible in the contaminated spectrum.}
	\label{fig:skyback}
\end{figure}

We rescale these files to represent the sky background at the LBT using measurements taken by Ref~\citenum{Pedani_14}. With an H-band magnitude of 14.29~mag/arcsec$^2$, the night sky flux density at the LBT is higher than Manua Kea by a factor of 36. Due to the small collecting area of iLocater's fiber, the addition of sky background, even in the presence of NIR OH emission lines, is negligible. Figure~\ref{fig:skyback} shows a direct comparison of sky background collected by a 42~mas SMF versus a 1'' MMF fiber. We find that sky background levels consistent with iLocater's on-sky collecting area introduce a 6 cm/s offset with no correction applied. 

\subsubsection[Neighboring Stars]{Neighboring Stars ($\sigma_{\rm stars}$)} 
Undetected neighboring stars (binaries, hierarchical triples, and unrelated sources) introduce additional absorption lines in measured spectra that can confuse the interpretation of RV measurements\cite{wright_13}. Compared to seeing-limited instruments, iLocater will have an improved ability to detect nearby companions to bright host stars by: (i) using high resolution spectroscopy to search for additional sets of absorption lines; and (ii) using AO imaging to identify other sources that may be in close angular proximity to the target. 

\subsection[Polarization Noise]{Polarization Noise ($\sigma_{\rm pol}$)}
\label{sec:polarization}
The single spatial mode in a SMF supports two orthogonal polarization modes $E_y$ and $E_x$. These modes are commonly referred to as S-polarization and P-polarization when referencing a physical surface in an optical system such as a diffraction grating. In an ideal fiber, these transverse waves travel with the same phase and group velocity, however, in practice, imperfections in the fiber core induce birefringence that changes the index of refraction in one axis relative to the other. As a consequence, the $E_x$ and $E_y$ components experience a delay that alters the output polarization state which manifests itself through a rotation of the electric field vector. The polarization state can change as a function of time when the fiber is subject to variable ambient conditions (changing temperature, bending, twisting, lateral stress, stretching, etc). In the presence of random perturbations, the fiber continuously and unpredictably modulates the magnitude of S- and P-polarization components. This causes the output ratio of S and P components to vary even with a constant/stabilized input. Such a modulation impacts RV precision when interacting with polarization sensitive optics. For example, if a diffraction grating is used after the fiber, this effect causes a variable blaze function, $b_{eff}$, which is unstable and differs from the scalar blaze function (i.e. measured with unpolarized light).

To quantify this effect we describe polarization of a given optical beam as the sum of a polarized and unpolarized fraction. Unpolarized light behaves in a known way and is modeled as a combination of 50\% S-polarization and 50\% P-polarization. As we do not have measured efficiency curves for all of iLocater's grating orders, we use characteristic wavelength and amplitude offsets between a few measured S and P curves to generate synthetic efficiency for each order. A new simulation is run for fractional polarization in steps of 10\% and fractions of light in the P mode in steps of 10\%. The RV uncertainty is calculated by cross correlating the result of each simulation with a mask, using the unpolarized case as an offset. We find that without deblazing or continuum normalizing, highly polarized sources (such as lasers) with a changing field angle can introduce RV uncertainties on the order of several meters per second, consistent with Ref~\citenum{halverson_15}. iLocater will use a super-continuum source (30\% polarization) for wavelength calibration, which introduces $\sim$~1m/s whereas starlight (1-10\%) ranges from $\approx$ 4-40~cm/s.

We find that the degree of polarization (DOP) may be reduced by a factor of 2-3 using the recombination of temporally incoherent light, through the use of a 2$\times$2 fiber splitter. These so-called ``recirculating delay lines" are well documented in applied optics literature and can be added in series to increase effectiveness \cite{Lutz_92,Shen_99}. With the correct experimental setup, complete depolarization (99$\%$) of light has been achieved \cite{Shen_98}. Tests undertaken for the development of iLocater are consistent with these results when using a linear polarized laser source, however, only a factor of two has been achieved thus far using the supercontinuum source ($\mbox{DOP}=30\%$). Depolarization of light immunizes the grating from variable birefringence in the fiber, translating to a decrease in RV uncertainty. Fiber splices and connections used in this method introduce photon losses only acceptable for calibration light.

\begin{figure}
	\centering
	\includegraphics[width = 6in]{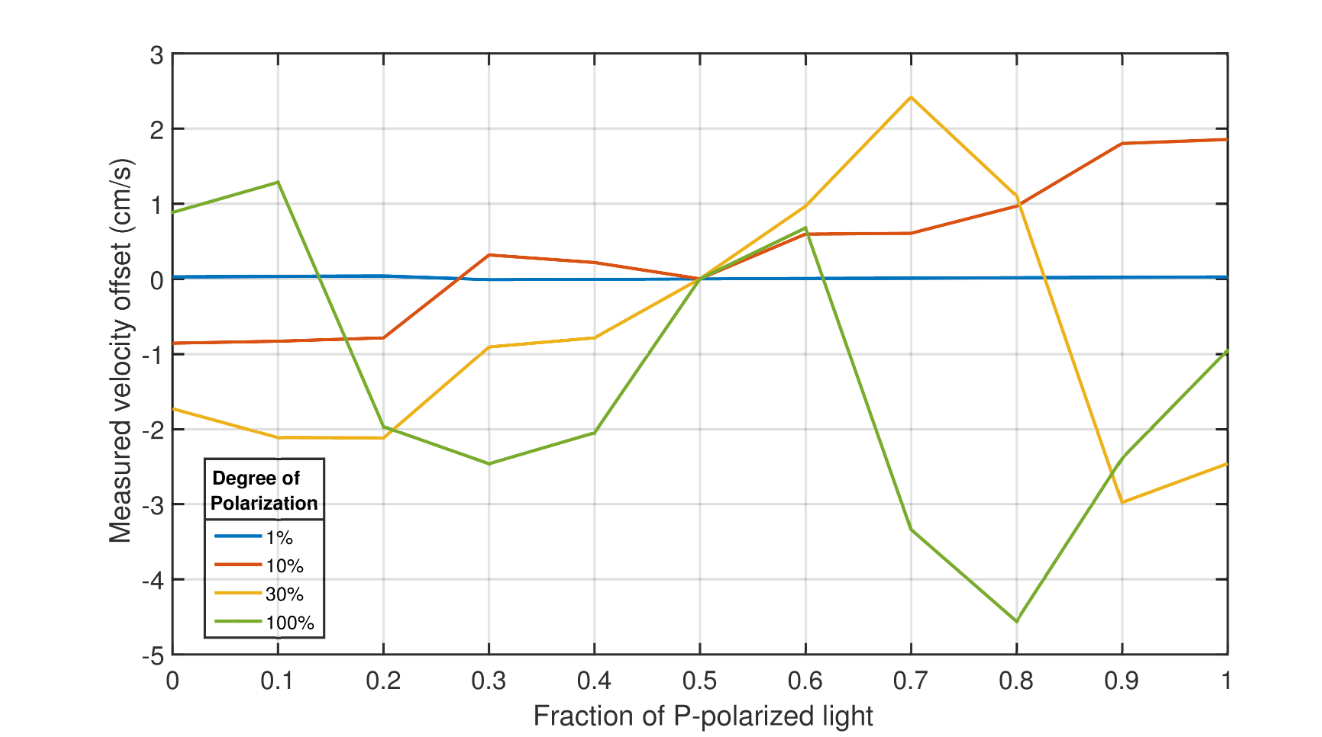}
	\caption{Measured velocity offset for various polarization states with software correction techniques. First, a measurement of the blaze function is used to remove the majority of order curvature, followed by a continuum normalization routine to flatten the remaining linear trends. Normalizing the spectrum reduces the error from polarization by two orders of magnitude.}
	\label{fig:Normalized}
\end{figure}

Removing the blaze function and applying a continuum normalization procedure to the spectral order prior to cross-correlation during the data reduction process reduces any remaining Doppler shifts caused by polarization effects to negligibly small levels. Figure~\ref{fig:Normalized} shows how blaze function removal and continuum normalization reduces errors to the noise floor of the iLocater RV pipeline. We find that 10\% polarized starlight has a maximum range of 3~cm/s and a more strongly polarized calibration source (30\%) has 5~cm/s. More detailed results are presented in (Bechter A. et al. 2018(b) in prep), including a characterization of the Stokes parameters through the PEPSI fiber feed at the LBT. These measurements are translated into variable fiber birefringence and propagated to the detector face with measured R6.1 polarization efficiency. Improved mitigation techniques are an on-going project. We allocate $\sigma_{\rm pol}$ = 10~cm/s until further testing is completed. 

\subsection[Thermo-mechanical]{Thermo-mechanical ($\sigma_{\rm thermal}$)}
\label{sec:thermal}

Thermo-mechanical RV uncertainties are based on material response for a given environmental change. Thermal and/or mechanical perturbations to the spectrograph optical components, mounts, and board in the dispersion direction translate into Doppler shifts at the detector focal plane. To quantify these effects, the instrument response is translated from a physical component into a pixel motion in the detector focal plane, and then calculated as a wavelength shift. The timescale for a perturbation is assumed to be less than a typical exposure, such that shifts in the dispersion direction manifest as untracked RVs. 

Thermal perturbations ($\Delta T$) alter the physical dimensions of material at operating temperature according to a specified coefficient of thermal expansion ($\alpha$). Precise Doppler instruments must be stabilized to the level of $\approx$1 milli-Kelvin or better, which is the requirement on the iLocater thermal control system. A uniform 1~mK thermal change is assumed for all materials in the following sections. This simplified approach is implemented to calculate an approximate RV uncertainty using analytic equations, and Zemax simulations. The following subsections outline the methods and results for each RV uncertainty term related to thermo-mechanical stability including: the Grating response in $\S$\ref{sec:grating}, optical board and optics deformation in $\S$\ref{sec:board}, pressure stability $\S$\ref{sec:pressure}, and vibrations stability in $\S$\ref{sec:vibrations}. 

\subsubsection[Gratings]{R6 Grating ($\sigma_{\rm gratings}$)}
\label{sec:grating}
The primary dispersing element is an R6.1 Echelle grating made from Zerodur and has 13.33 grooves/mm at room temperature. At an operating temperature of 58 K, $\alpha$ for Zerodur is $-0.15 \times 10^{-6} K^{-1}$. The relationship between the grating size, L, and wavelength, $\lambda$:
\begin{equation}
\frac{\lambda}{d\lambda} = \frac{L}{dL},
\end{equation}
can be derived from the well known grating equation. Assuming a 1 mK perturbation, and the expansion/ contraction of the grating follows $dL = L\alpha\Delta T$, the RV uncertainty can be calculated from:
\begin{equation}
\Delta RV = \alpha\Delta T~c = 4.5~\mathrm{cm/s}.
\end{equation}
We hold 5~cm/s as the contribution from the Zerodur R6 grating irrespective of the optical board material selection.


\subsubsection[Optical Board and Optics]{Optical Board and Optics
($\sigma_{\rm optics}$)}
\label{sec:board}

The combined thermal effect of the optical board and optics is measured in Zemax by perturbing the system by 1 mK. In this section we include a comparison of aluminum and Invar/Zerodur designs to highlight the importance of material selection. While thermal gradients are likely to be smaller for an aluminum system when compared to Invar/Zerodur due to the relatively high conductivity of aluminum. Considering this, the 1~mK uniform step is potentially more realistic for an aluminum system. Mounts are not included in this analysis.  

\begin{figure}
	\centering
	\includegraphics[width = 6in]{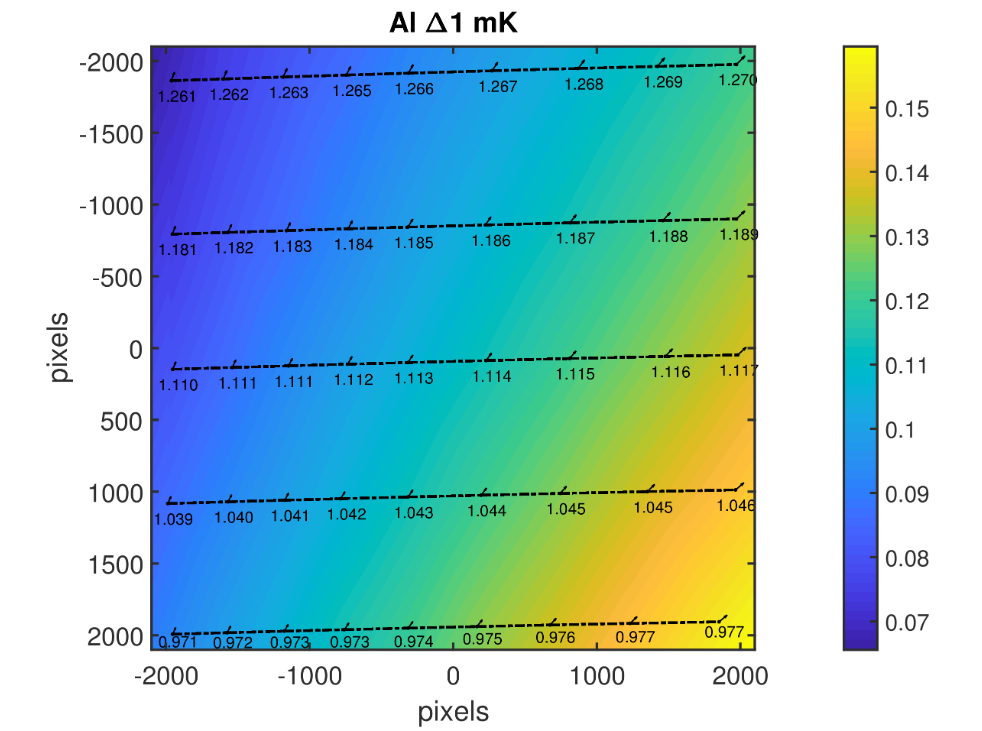}	
	\caption{RV uncertainty map evaluated at the H4RG detector focal plane for a 1 mK change in temperature at a nominal operating temperature of $T=60$ K for the aluminum design. The RV uncertainty is calculated for five orders shown as dashed black lines. The magnitude and \emph{direction} of centroids are shown using arrows. RV uncertainties are calculated as the centroid motion in the dispersion direction, the magnitude of which is shown on the color-scale.}
	\label{fig:ThermalRV}
\end{figure}

We start with $\alpha$ for Zerodur as $\mbox{CTE}=-0.15\times10^{-6} K^{-1}$ and aluminum $\mbox{CTE}=17.5\times10^{-6} K^{-1}$. The centroids for nine wavelength samples across five orders are tracked from nominal temperature to 1 mK above. From these centroids we calculate a pixel shift, $\Delta x$ (change in dispersion direction), then $\Delta\lambda$ using the wavelength solution, and finally $\Delta v$. Figure \ref{fig:ThermalRV} shows the results of this calculation using aluminum. All orders have a preference to shift diagonally as shown by the arrows on each wavelength sample. At no point on the detector is there zero shift in the dispersion direction. The maximum and minimum RV values on the detector face are 20 cm/s and 7 cm/s. 

\begin{figure}
	\centering
	\includegraphics[width = 6in]{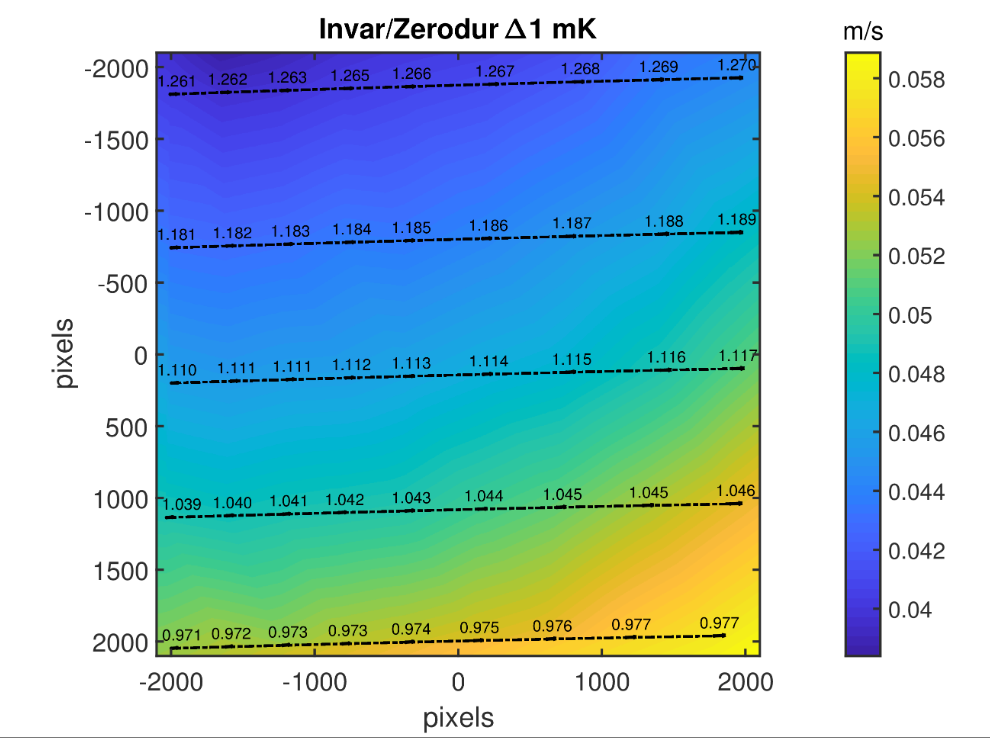}	
	\caption[RV uncertainty map for an Invar spectrograph.]{RV uncertainty map evaluated at the H4RG detector focal plane for a 1 mK change in temperature at a nominal operating temperature of $T=60$ K for the Invar/Zerodur design. The RV uncertainty is calculated for five orders shown as dashed black lines. The magnitude and \emph{direction} of centroids are shown using arrows. RV uncertainties are calculated as the centroid motion in the dispersion direction, the magnitude of which is shown on the color-scale.}
	\label{fig:ThermalRVInvar}
\end{figure}

We next compare the aluminum design to an Invar/Zerodur design. In this case, the space between optics has a CTE value of $\alpha$ = $0.73\times10^{-6} K^{-1}$ whereas optics take on the CTE of Schott Zerodur. Figure \ref{fig:ThermalRVInvar} shows the results of this calculation. Interestingly, all orders have a preference to shift in the dispersion direction, making the vector arrows at each wavelength sample difficult to distinguish. Similarly to the aluminum system, at no point on the detector is there a zero shift in the dispersion direction. The max and min RV values however are much smaller than the aluminum system at 5.7 cm/s, and 3.9 cm/s. Furthermore, the Invar/Zerodur design not only has a smaller RV offset, but also a more uniform shift across the detector face. As the max value is well below the allotted value of 10~cm/s, it is possible that no additional thermal correction would be required to account for thermal drifts during a 30-60 minute observation. 

As a result from this study, we assign a (peak-to-valley) value of $\sigma_{\rm thermal}=20$ cm/s for the worst case uncorrected RV shift until further simulations are conducted which account for the effects of optical mounts. We conservatively assume a correction of 50\% can be achieved on thermo-mechanical terms through calibration data acquired during observations (75\% is assumed in Ref~\citenum{halverson_16}). 

\subsubsection[Pressure Stability]{Pressure Stability ($\sigma_{\rm pres}$)} 
\label{sec:pressure}
Pressure variations within the spectrograph will lead to changes in refractive index. As the refractive index is altered, the wavelength of light propagating through the spectrograph will also change, which can also be understood as an RV uncertainty. Following the framework of Ref~\citenum{stefansson_16}, we estimate the shift in wavelength due to pressure changes using: 
\begin{equation}
\lambda = \frac{\lambda_{vac}}{n(P,T)},  
\end{equation}
where $\lambda_{vac}$ is the wavelength in vacuum, and $n(P,T)$ is the index of refraction as a function of temperature and pressure given by Ref~\citenum{Edlen_66}. RV uncertainties are calculated using the Doppler equation: 
\begin{equation}
\Delta v = c \frac{\Delta \lambda}{\lambda_{vac}}.
\end{equation} 
Using a representative operating temperature of $T=60$ K and (pessimistic) change in pressure of $\Delta P = 10^{-6}$ Torr (short term fluctuations may be closer to $10^{-8}$ Torr), we calculate an RV error of $\sigma_{\rm pres}=0.05$ cm/s. Relative to other terms in the error budget, we conclude pressure fluctuation induced RV uncertainties are completely negligible at variations of $\Delta P \leq 10^{-6}$ Torr .

\subsubsection[Vibrations]{Vibrations ($\sigma_{\rm vib}$)}
\label{sec:vibrations}
Mechanical and acoustic vibrations effectively smear out the spectrograph response creating a wider line-spread function which degrades resolution and thus RV measurement precision. We require that vibrations contribute negligibly to iLocater's RV error budget. In order to minimize vibrations and other disturbances from temperature and pressure changes, the spectrograph will be located far from the telescope's immediate environment.

We require that vibrations contribute negligibly to iLocater's spectral resolution. The reduction in spectral resolution, $\Delta R$, must satisfy:
\begin{equation}
\Delta R / R < 0.01
\end{equation}
The iLocater line spread function will subtend 3 pixels on an H4RG infrared array. If
each pixel is 10 $\mu$m wide, then we require that vibration amplitudes, $A$, be smaller than
\begin{equation}
A < \Delta R / R (10) (3.0) = 0.3 \; \mu m.
\end{equation}
Notice that vibrations in the direction of the optical axis where the beam is focused onto the infrared array are of little concern since the spectrograph camera optics have a depth of focus (DOF) of approximately:
\begin{equation}
\mbox{DOF} = 2 f_{\#}^2 \lambda \approx 2 (16)^2 (1.0) = 32 \; \mu m
\end{equation}
Ideally, a mounting design should preferentially isolate vibrations in this direction. Excitation of instrument resonances are also an important consideration. Assuming a mechanical damping system will be employed to relax vibrations by a factor of $10\times$, a cryocooler that produces residual vibrations with amplitude of 3~$\mu$m may be accommodated. Suppression levels as high as two orders of magnitude have been demonstrated elsewhere for space applications (R. G. Ross, Jet Propulsion Laboratory, SPIE, 2003, ``Vibration Suppression of Advanced Space Cryocoolers -- An Overview").

\begin{figure}[h]
	\includegraphics[width=\textwidth]{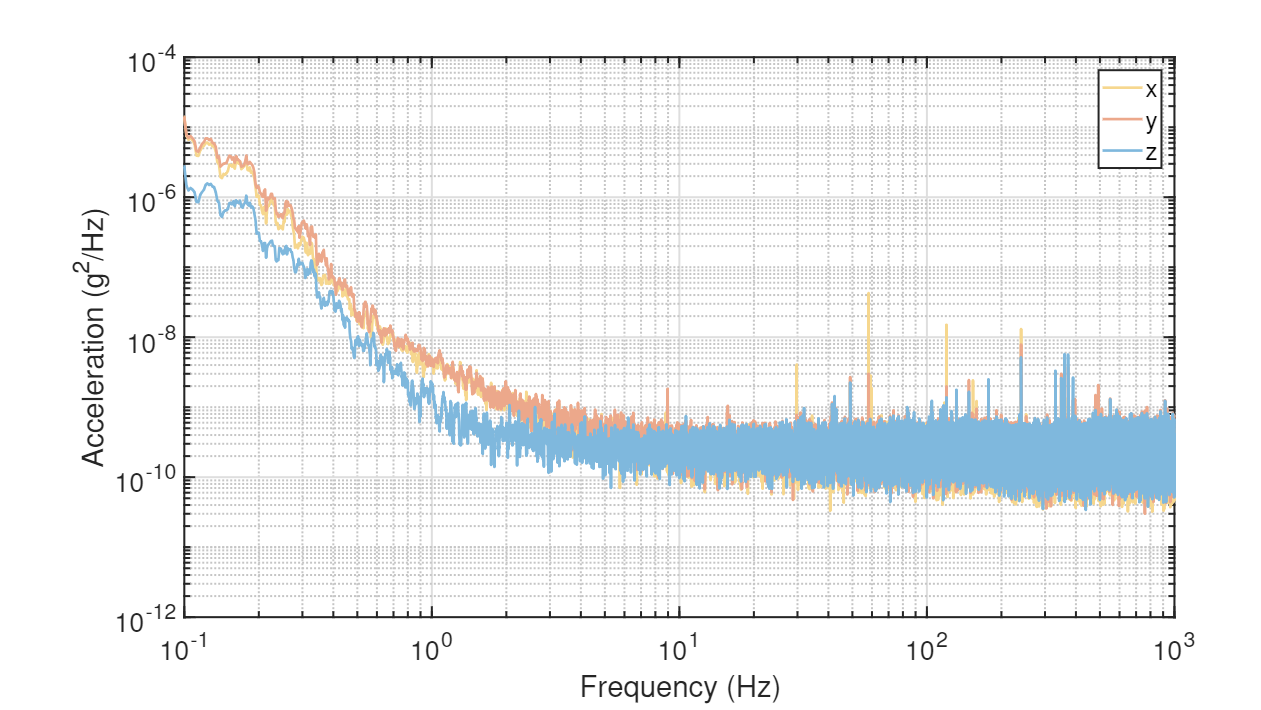}
	\caption{Power spectral density of the mounting location for iLocater. The x and z axes are in the plane of the optical board and the y axis is perpendicular.}
	\label{fig:vib_spec} 
\end{figure}

The spectrograph mounting location is remote from the telescope and has been shown from accelerometer measurements to be very isolated with only a few vibrational resonances at 30, 60, 120 and 240 Hz as shown in Figure \ref{fig:vib_spec}. Translating the acceleration PSD into a displacement PSD can be achieved by dividing by $\omega^4 = (2\pi f)^4$ and filtering using a high pass filter to remove accelerometer noise below 1Hz. Using this method we estimate the rms value of non-dampened displacement to be a few microns. As with the cryocooler requirements above, this requires mechanical isolation of about an order of magnitude.  

\subsection[Wavelength Calibration]{Wavelength Calibration ($\sigma_{\rm cal}$)}
\label{sec:wavecal}
All Doppler measurements are benchmarked against an incredibly precise and accurate wavelength calibration unit. iLocater will use a temperature stabilized, Fabry-P\'erot etalon. The intrinsic stability of of the etalon has been demonstrated at the 3 cm/s level from calibration based on the hyper-fine lines of a Rubidium gas cell \cite{gurevich_14,schwab_15}. 

\begin{figure}
	\centering
	\includegraphics[width = 4in]{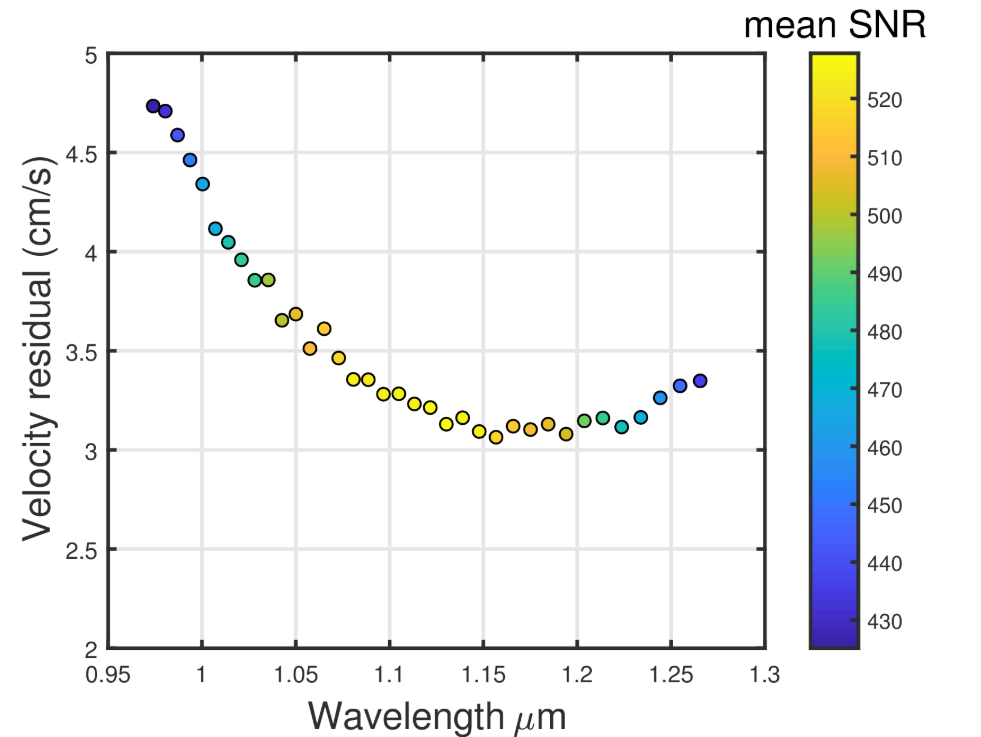}	
	\caption[Etalon Photon noise limited Doppler information]{Photon noise limited Doppler information for iLocater's etalon as a function of wavelength Each of the 36 order's performance is primarily dictated by the instrument throughput, with central orders able to achieve higher SNR than the top and bottom orders.}
	\label{fig:EtalonDop}
\end{figure}

In addition to intrinsic stability, the photon noise limited RV precision of the etalon lines contribute to the overall wavelength calibration uncertainty. We simulate the etalon transmission function using the prescribed free spectral range (FSR) and finesse to calculate the RV precision per spectral order shown in Figure~\ref{fig:EtalonDop}. The average per-order information is 3.5 cm/s with a SNR of $\sim$400. Combining information from all 36 orders results in a final photon noise value of 1 cm/s. We conservatively allocate $\sigma_{\rm cal}=5$ cm/s for the combined effects of intrinsic etalon stability and RV information available on iLocater's detector. 

\subsection[Tellurics]{Tellurics ($\sigma_{\rm tell}$)}
\label{sec:tellurics}
Changes in the column density of Earth's atmosphere modify the stellar spectrum in time creating systematic errors. These telluric absorption lines permeate the NIR regions of our atmosphere and ``bounce around" at hundreds of meters per second \cite{bean_10}. Most pronounced in the JHK bands, tellurics have limited all previous attempts at precise NIR Doppler measurements. Efforts to circumvent this problem generally involve wholesale removal of such features using software, however this approach severely limits the amount of RV information extracted from remaining starlight. For example, in the K-band telluric absorption results in an 80\% loss of spectral data \cite{reiners_10}. Fortunately, there exists a NIR band relatively free of atmospheric absorption. The Y-band is a region spanning the $\lambda=0.97-1.12 \: \mu$m range that is clean compared to other infrared windows \cite{hillenbrand_13}; it also provides access to deep stellar absorption lines rich in Doppler information especially for M-dwarfs \cite{kirkpatrick_99}. 

\begin{figure}
	\centering
	\includegraphics[width = 6.5in]{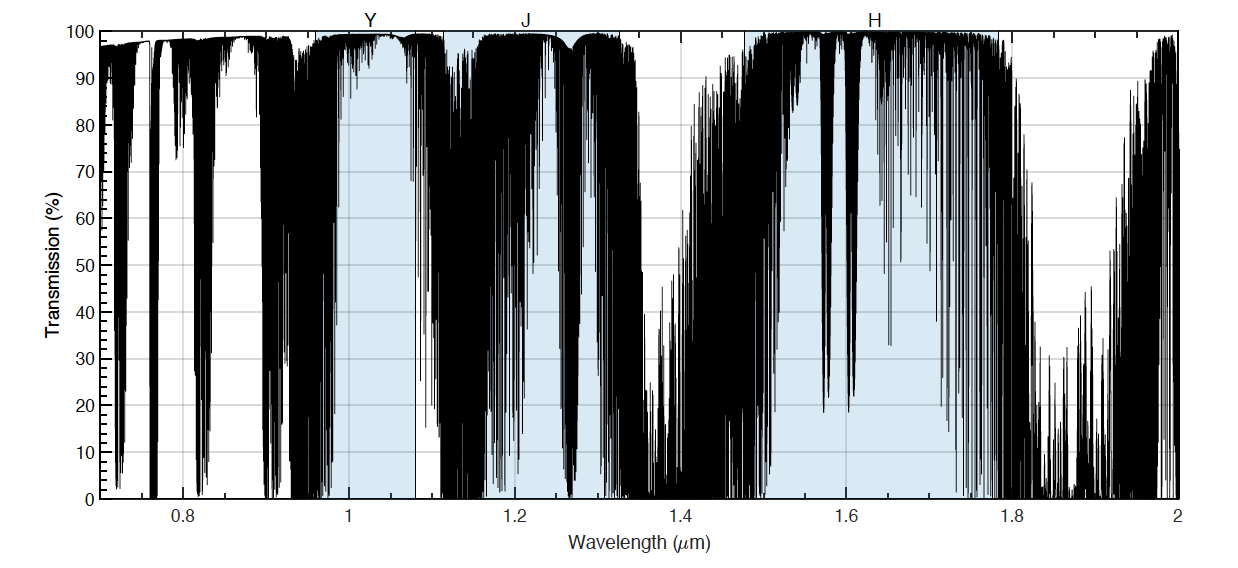}	
	\caption{Telluric contamination in the shaded Y- and J-bands is $\sim$ 20-50$\%$.}
	\label{fig:tellurics}
\end{figure}

Telluric contamination is reported to be 20\% in the Y-band and 55\% in the J-band with a considerable blanketing of lines between the two bands ($\lambda=1100-1160$ nm) \cite{reiners_10}. Figure~\ref{fig:tellurics} shows an example of atmospheric transmission across the iLocater bandpass. The reduction in Doppler content results in higher SNR requirements in order to acheive the same RV precision. This increase is reflected in photon noise requirements derived in $\S$\ref{sec:photon_noise}.  

Short of having an optimized removal strategy for telluric lines, we asses the uncorrected value by first cross correlating a telluric free spectrum with an example contaminated spectrum multiplied by a telluric spectrum. This results in an RV error of $\sim100$~m/s. If telluric lines aren't removed or corrected, the resulting RV error completely overwhelms the error budget. This emphasizes the importance of having a comprehensive telluric mitigation strategy. Telluric calibration is currently the most uncertain term in iLocater's RV error budget and we allocate $\sigma_{\rm tell} \approx 20$ cm s$^{-1}$ to this effect. Careful treatment of telluric absorption lines will be critical for EPRV instrument to achieve precision $<$ 1m/s \cite{seifahrt_10}. 

\subsection[Barycentric Correction]{Barycentric Correction ($\sigma_{\rm BC}$)}
\label{sec:bary}
Earth's motion around the Sun creates a systematic Doppler effect of order $\pm 30$ km s$^{-1}$, the removal of which must make a negligible contribution to iLocater's error budget. iLocater will use a publicly available barycentric correction code (BARYCORR) written in IDL that corrects for Doppler errors introduced by the spin and orbit of the Earth at the level of several cm s$^{-1}$ \cite{wright_14}. Originally developed for the California Planet Search, the BARYCORR algorithm uses a fully general relativistic treatment of space motions and has been tested against the best pulsar timing package, TEMPO2 \cite{hobbs_06,edwards_06}. 

\subsection[Pipeline]{Pipeline ($\sigma_{\rm software}$)}
\label{sec:software}
Measuring Doppler shifts at the level of a tiny fraction ($10^{-4}$) of a pixel requires an exquisitely precise software pipeline. We have developed a custom, end-to-end data analysis program that uses a combination of Matlab, Python, and Zemax to model the spectrograph signal as incident onto an infrared detector. The program accurately captures the optical response of the instrument based on the actual spectrograph design. Additionally, physical effects that can impact RV precision (hot pixels, dark current, read noise, instrument stability, the wavelength calibration unit, barycentric motion, telluric absorption, OH emission, etc.) are incorporated into the program to mimic conditions expected to be encountered at the observatory. 

There are small algorithmic uncertainties inherent in every data reduction pipeline that have the potential to affect RV precision. Some of these errors include: interpolation of bad detector pixels, cosmic-ray identification and removal, spectral order tracing shifts, etc. It is particularly difficult to quantify their individual effects on RV precision as they are predominantly random uncertainties. Therefore, to estimate the magnitude of these errors, we simulate many noiseless data frames, varying only stellar RV, and measure the residual error in the pipeline's recovery of the injected RV. Our results show that none of the residual software errors impart an RV uncertainty greater than 5 cm/s. Because this number does not likely encompass all software error sources, we reserve a value of 10 cm/s for software algorithms, which is in agreement with Ref~\citenum{halverson_16} and Ref~\citenum{terrien_14}.   


\subsection[Detector Noise]{Detector Noise ($\sigma_{\rm H4RG}$)}
\label{sec:detector}

The 4k$\times$4k, hybrid structured mercury-cadmium-telluride H4RG-10 is a new NIR detector, originally developed as part of NASA's WFIRST mission. While these detectors provide outstanding quantum efficiency ($\sim$90\%), they have not yet been thoroughly tested in ground-based RV spectrographs, which presents some inherent risks for the project. H4RG detector characteristics overlap substantially with well known CCD noise types, but also have some unique noise attributes, including: alternating column noise (ACN), picture frame noise, persistence, inter-pixel capacitance (IPC), and random telegraph noise.     

We investigate these effects on RV precision using our spectrograph simulator and RV pipeline in combination with the ``HxRG Noise Generator,'' from Ref~\citenum{rauscher_15} which emulates the effects H4RG read noise, broken down into the components: white noise, correlated and uncorrelated pink (1/f) noise, ACN, and picture frame noise. Adding to this, we developed our own simulations of persistence and dark current. The details of these simulations are outlined in Bechter E. et al. 2018 (b), in prep. The summary of results for this study are shown in Table~\ref{tab:errorbudget}. White noise, correlated \& uncorrelated pink noise, ACN, picture frame noise, and dark current all contribute only a few cm/s to the RV error budget. However, depending on the previous target observed and time between observations, persistence can be potentially overwhelming. The best current mitigating strategies for persistence are to plan observations throughout each night such that persistence is minimized from one exposure to another\cite{Artigau_18}. It should also be noted that these simulations do not yet encompass every source of detector noise such as IPC, random telegraph noise, and defects (hot/dead pixels). As each manufactured detector can vary significantly in its performance and noise characteristics, it is difficult to assign a single value to the RV error budget without first characterizing the detector in a lab setting. For now, we assign a conservative 30 cm/s for detector related RV uncertainties. 

\section{Summary and Future Work}
\label{sec:conclusion}
As RV precision goals are pushed to well below 1 m/s, simulating instrument performance and constructing a reliable error budget becomes an essential aspect of the design and iteration process. We extend the approach used by Halverson et al. 2016 for multi-mode fibers to a SMF-fed spectrograph using iLocater a design reference.

A summary of the error budget is shown in Table~\ref{tab:errorbudget}. The first and second columns correspond to the same categories and terms shown in the box diagram in Figure~\ref{errbud}. Each term lists an `Uncorrected' value that represents RV uncertainty with no correction applied. These are taken as baseline values and should identify the key terms to study calibration and mitigation methods. Corrected values are shown for terms that have a well understood or assumed correction (at the time of writing). Allocated values represent budgeted values for entire categories taking the previous two into account.

It is clear that photon noise is a significant contributor to the overall error budget. The faintest of targets we simulate at I=10, contribute 60cm/s, while the brightest at I=4, 20cm/s. Using a SMF offers an appealing method to increase spectral resolution without significantly increasing the instrument footprint. The cost of such a change however is an increase in photon noise through significant injection losses at the SMF. Photon noise is perhaps the most difficult term for a single-mode EPRV instrument to achieve. Improved grating efficiency and WFS upgrades are the most readily available method for improving this term. We note that diffraction-limited designs for existing telescopes could be identically ``cloned" for ELTs. 


Calibration of known systematics is also crucial for diffraction-limited spectrographs to meet RV error budget objective. Several terms warrant a comprehensive study on their own (e.g. entire PhD theses). For example, tellurics pose the biggest known systematic error without a clear near-term strategy for correcting.\cite{plavchan_18} Polarization and thermo-mechanical effects make non-negligible contributions but have more clear mitigation strategies. Harmonizing material properties used for optics, opto-mechanical components, and detectors is an integral part of improving RV precision, particularly in the NIR. Finally, detector noise terms (such as persistence) require further theoretical and laboratory investigation.  



\begin{sidewaystable*}[p]
	\footnotesize
	\begin{threeparttable}
		\centerline{
			\begin{tabular}{|c|l|ccc|c|l|}
				\hhline{|=======|}
				\multicolumn{7}{|c|}{Error Budget Summary (cm/s)} \\
				\hhline{|=======|}
				Category 			& Terms  				&Uncorrected 	& Corrected 	& Allocated & Currently in Budget	& Driving factor\\
				\hline
   SMF Injection& AO \& ADC		 		& 6				&$\dagger$	 	& 6		    &$\checkmark$ 	& Strehl ratio \& dispersion\\
				& Alignment		 		& 8			 	&$\dagger$	 	& 8		    &$\checkmark$ 	& Tip\& Tilt residuals \\
				& Modal noise	 		& 0			 	&-			 	& 0			& 				& Fiber diameter \\
				& Cladding modes 		& 0			 	&$\dagger$	 	& 0			&				& Mode matching \\
				& Deformable mirror		& 1				&- 				& 1		    &$\checkmark$ 	& Actuator amplitude/frequency\\
				& \textbf{Subtotal}		& \textbf{10}	&			 	&\textbf{10}&$\checkmark$   & -\\
				\hline
Projected Sky Angle	& Sky background	& 6 		 	&$\dagger$	 	& 10   		&$\checkmark$	& OH emission\\
				& Neighboring stars	    & - 		 	&-			 	& 0			&$\checkmark$	& Stellar Contamination\\
				& \textbf{Subtotal}		& \textbf{6}	&			 	&\textbf{10}&$\checkmark$ 	& -\\
				\hline
Polarization	& Calibration			& 100 		 	&5			 	& 9  		&$\sim$     	& Calibration source\\		
				& Stellar				& 40 		 	&3  		 	& 4  		&$\sim$	        & R6.1 S\&P \\
				& \textbf{Subtotal}		& \textbf{108}	&\textbf{5.8}	&\textbf{10}&$\sim$ 	    & \\
				\hline
Thermo-mechanical& R6 Grating			& 4.5 		 	&2.3			& 3	 		& $\checkmark$	& Zerodur class\\
				& Optics/Board			& 20 		 	&10				& 10		& $\sim$	    & Material CTE (aluminum) \\
				& Vibrations			& -			 	&1				& 1		    & $\checkmark$	& Cryo coolers\\
				& Pressure				& 0			 	&0				& 1			& $\checkmark$	& Refractive index change\\
				& \textbf{Subtotal}		& \textbf{21}	&\textbf{10.3}	&\textbf{10}& $\sim$		& -\\
				\hline
Calibration 	& Stability				& 3			 	&3				& 4			& $\checkmark$	& Rb triplet, mechanical stability\\
				& Photon Noise			& 1			 	&1				& 3			& $\checkmark$	& Doppler information\\
				& \textbf{Subtotal}		& \textbf{3.2}	&\textbf{3.2}	&\textbf{5}	& $\checkmark$	& - \\
				\hline				
Reduction Pipeline & Algorithms			& 5			 	&$\dagger$		& 10		& $\checkmark$	& - \\
				\hline
Detector effects& White noise           & 5-10	        &$\dagger$		& 10  		& $\checkmark$	& Johnson noise in pixel interconnects \\
				& Correlated Pink noise	& 8-10		 	&$\dagger$		& 10		& $\checkmark$	& Output field effect transistor \\
				& Uncorrelated pink noise& 10	 	 	&$\dagger$		& 10		& $\checkmark$	& - \\
				& Alternating column Noise& 10		 	&$\dagger$		& 10 		& $\checkmark$	& Voltages multiplexing \\
				& Picture frame noise	& 1-2		    &$\dagger$		&  2		& $\checkmark$	& Readout integrated circuit \\
				& Dark Current	        & 3 		    &$\dagger$		& 10 		& $\checkmark$	& Op. temperature \\
				& Persistence	        & 0-1000		&$\dagger$		& 20		& $\times$		& Observational strategy \\
				& \textbf{Subtotal}		&\textbf{17-1000}&\textbf{17}	&\textbf{30}& $\sim$        & Presistence correction\\
				\hline
External		& Tellurics				& $\sim$1000  	&$\dagger$ 		& 20		&$\times$		& Correction method development \\
				& Barycentric Velocity	& $\sim$3$\times10^{6}$	&$\dagger$ & 3 		&$\checkmark$	& Exposure meter \\
				\hline
Totals 			& Instrument Systematics& 111-1000		 &25			& 36		&$\checkmark$	& H4RG Detector \\	
				& External Systematics	& $\sim$3$\times10^{6}$ &$\dagger$&20.2	    &$\sim$			& Tellurics \\
				& Photon Noise 			& 20-60		 	&20-60			&60		    &$\checkmark$	& Strehl ratio, R6 efficiency\\
				\hline
				& Total 			    & -		 	    & -				&73		    &           	& With telluric/persistence correction\\
				\hhline{|=======|}
		\end{tabular}}
		\caption{Summary of RV budget error terms. Each term is listed with an `Uncorrected', `Corrected' (if any is expected) and `Allocated' values. Terms with a dash (-) indicate they have not been explicitly quantified as opposed to zeros which indicate a negligible value of $\le$ 0.5 cm/s. Daggers ($\dagger$) indicate some correction may be possible but has not been quantified. The `Currently in Budget' symbols represent whether individual terms and categories are within the allocated values ($\checkmark$), close to allocated values but require corrective methods ($\sim$), and those not within the budget ($\times$)} 
		\label{tab:errorbudget}
	\end{threeparttable}
\end{sidewaystable*}


 

\acknowledgments 
JRC acknowledges support from the NASA Early Career and NSF CAREER Fellowship programs. 

\bibliography{report} 
\bibliographystyle{spiebib} 

\end{document}